# Local Structure Dictates Ionic Transport and Mechanical Properties in Glassy Solid Electrolytes for Lithium Batteries

Yong Li [a], Tao Du [b], Rasmus Christensen [a,c], Timothée Jamin [a], Zhencai Li [a], Qi Zhang [d], Xiaoyi Xu [a], Kasper Tolborg [a], Yuanzheng Yue [a], Morten M. Smedskjaer [a,*]

[a] *Department of Chemistry and Bioscience, Aalborg University, 9220 Aalborg East, Denmark*
[b] *Department of Applied Physics, The Hong Kong Polytechnic University, Kowloon, Hong Kong 999077, China*
[c] *Department of Applied Physics, Tohoku University, Sendai, Japan*
[d] *Corning Research Center in China, Corning Incorporated, Shanghai, China*
[*] *Corresponding author. E-mail: mos@bio.aau.dk*

**Abstract**

Electrolytes composed of sulfide and halide glasses are promising candidates for all-solid-state lithium batteries owing to their processability, lack of grain boundaries, and relatively high ionic conductivity. Nevertheless, their ionic conductivity and mechanical properties are still not satisfying for the real-world applications. Significant advances in solid electrolytes require a thorough understanding of their microstructures. Here, we reveal the connections among structure, ionic transport properties, and mechanical stability in a series of glassy solid electrolytes by employing molecular dynamics simulations based on a machine learning interatomic potential. Specifically, we explore how the interplay between B-S and P-S networks in glassy Li-S-P-B-I (LSPBI) governs ionic conductivity and deformation behavior. The introduction of $P_2S_5$ into a $B_2S_3$-based glass induces a critical structural transformation, through which both ionic conductivity and mechanical nano-ductility can be enhanced. For a moderate $P_2S_5$ content, incorporated $PS_4$ units depolymerize the rigid boron framework, creating percolative diffusion pathways for fast ionic transport. Concurrently, the flexible P-S-P configurations enable energy dissipation through bond bending, leading to the brittle-to-ductile transition. However, excessive $P_2S_5$ increases the fraction of polyphosphates (e.g., $P_2S_6$ and $P_2S_7$), thereby polymerizing the structural network and ultimately impeding $Li^+$ mobility. Our work thus provides atomistic principles for engineering glass electrolytes with balanced ionic conductivity and mechanical robustness.

## 1. INTRODUCTION

The path to a sustainable energy future relies on batteries that overcome the trade-off between energy density, safety, and lifespan. [1, 2] Lithium-ion batteries using liquid organic electrolytes face flammability and electrochemical instability issues, limiting their voltage windows.[3, 4] All-solid-state batteries containing a solid-state ionic conductor offer a promising path to mitigate these risks while enabling higher energy densities.[5, 6] Among solid-state electrolyte (SSE) candidates, including ceramics[7] and polymers[8], inorganic glassy electrolytes[9] hold distinct promise due to their isotropic ionic conduction, absence of grain-boundary resistance, and excellent compositional flexibility. Oxide-based glasses (e.g., Li-B-P-O) offer chemical stability and a wide electrochemical window but moderate conductivity.[10, 11] In contrast, sulfide glasses (e.g., Li-P-S) achieve superior $Li^+$ conductivity due to the high polarizability of sulfur anions, yet their practical application is limited by poor chemical stability against moisture and narrow electrochemical windows.[12]

This limitation calls for the development of glass materials combining the high conductivity of sulfides with enhanced stability. A promising strategy is the mixed-network-former approach, particularly combining thioborate ($B_2S_3$) and thiophosphate ($P_2S_5$) networks.[13, 14] Extending this, introducing polarizable halide anions (e.g., $Cl^-$, $Br^-$, $I^-$) creates mixed-anion systems, making halide SSEs a promising candidate class.[15] A recent experimental study[16] identified lithium thioborophosphate iodide (LSPBI) glasses as a compelling multi-anion system, where sulfide (high conductivity), boron/phosphate (stability), and iodide (polarizability) synergistically yield high $Li^+$ diffusivity. Atomistic simulations have also been used to explore their diffusion mechanisms.[17] However, while ionic conductivity is critical, practical deployment equally hinges on mechanical reliability.[18-20] During cycling, the solid electrolyte must withstand repeated stress from electrode volume changes without fracturing; cracking increases interfacial resistance and creates dendrite propagation pathways.[21, 22] Compared to well-investigated diffusion, the mechanical performance of multi-anion glasses like LSPBI, along with its atomic origins, remains unexplored. In network glasses, mechanical behavior is governed by competition between bond breaking and localized shear flow, linked to network connectivity and bond switching.[23,

[24] How the unique multi-anion environment of LSPBI glasses influences their deformation micro-mechanisms remains an open question.

At the atomic scale, accurate description of materials properties relies on reliable molecular dynamics (MD) simulations. The advent of machine learning interatomic potentials (MLIPs) has revolutionized this field by bridging the accuracy of density functional theory (DFT) with the time and length scales of classical molecular dynamics. Frameworks such as the DeePMD-kit[25] have been instrumental in this shift, and its latest iteration (DeePMD-kit v3)[26] incorporates hybrid descriptor models and improved training efficiency, making it a widely adopted platform for high-fidelity simulations of complex systems. Concurrently, the field has seen rapid architectural innovations, including MACE[27], which employs high-body-order equivariant message passing, and Allegro[28] which achieves massive parallelization through strictly local equivariant tensor products without message passing. Collectively, these MLIPs, trained on *ab initio* data, now enable unprecedented simulations of multi-component glasses such as LSPBI, providing the essential computational tool to uncover relationships between atomic-scale structure, ion transport, and deformation mechanisms.

In this work, we employ MD simulations, validated based on existing experimental structural and diffusivity data,[16] to understand the coupled mechanical and Li-ion transport properties of LSPBI glasses. Specifically, we focus on a series of LSPBI glassy electrolytes with the general formula $30Li_2S$ - $25B_2S_3$ - $45LiI$ - $aP_2S_5$, where $a$ ranges from 0 to 100 ($a$ = 0, 5, 10, 25, 50, 75, 100). Individual compositions are denoted as LSPBI-$a$X (e.g., LSPBI-$a$0, LSPBI-$a$5, …, LSPBI-$a$100). Our primary objectives are to: (i) characterize the composition-dependent lithium-ion transport, quantifying how the addition of $P_2S_5$ modulates diffusion mechanisms, activation barriers, and resulting ionic conductivity; (ii) characterize the composition-dependent fracture resistance and elucidate the atomic-scale mechanisms governing an observed brittle-to-ductile transition; and (iii) establish structure-property relations that link composition-induced topological changes to both the long-range $Li^+$ mobility and the propensity for inelastic mechanical deformation. By doing so, we aim to understand how the multi-anion chemistry of LSPBI can be strategically tailored to achieve the dual, and often competing, requisites of high ionic conductivity and superior mechanical resilience.[29] Overall, the study thus provides fundamental insights into the structure-ionic conductivity-mechanical performance relationships and design principles for engineering robust, high-performance glassy electrolytes for durable all-solid-state batteries.

## 2. RESULTS AND DISCUSSION

### 2.1 Training of MLIP for glass structure generation

To enable long time- and length-scale simulations of LSPBI glasses, we have here developed a dedicated DeePMD-kit MLIP trained on DFT/AIMD data and benchmarked it against AIMD and universal MACE potentials. Because experimental structural data are limited, AIMD simulations of LSPBI-*a*5 were used as the primary reference baseline. Transferability was further supported by the agreement between simulated and reported neutron-weighted S(Q) for $75Li_2S$ - $25P_2S_5$ (Supporting Figure S1). [30,31] The LSPBI glass sample was prepared via a melt-quench protocol. The density obtained from the AIMD simulation is approximately 2.28 $g/cm^3$, which is consistent with the experimentally measured value of 2.3 $g/cm^3$.[16] The equilibrated atomic structure of the LSPBI-*a*5 glassy electrolyte is shown in Figure 1*a*.

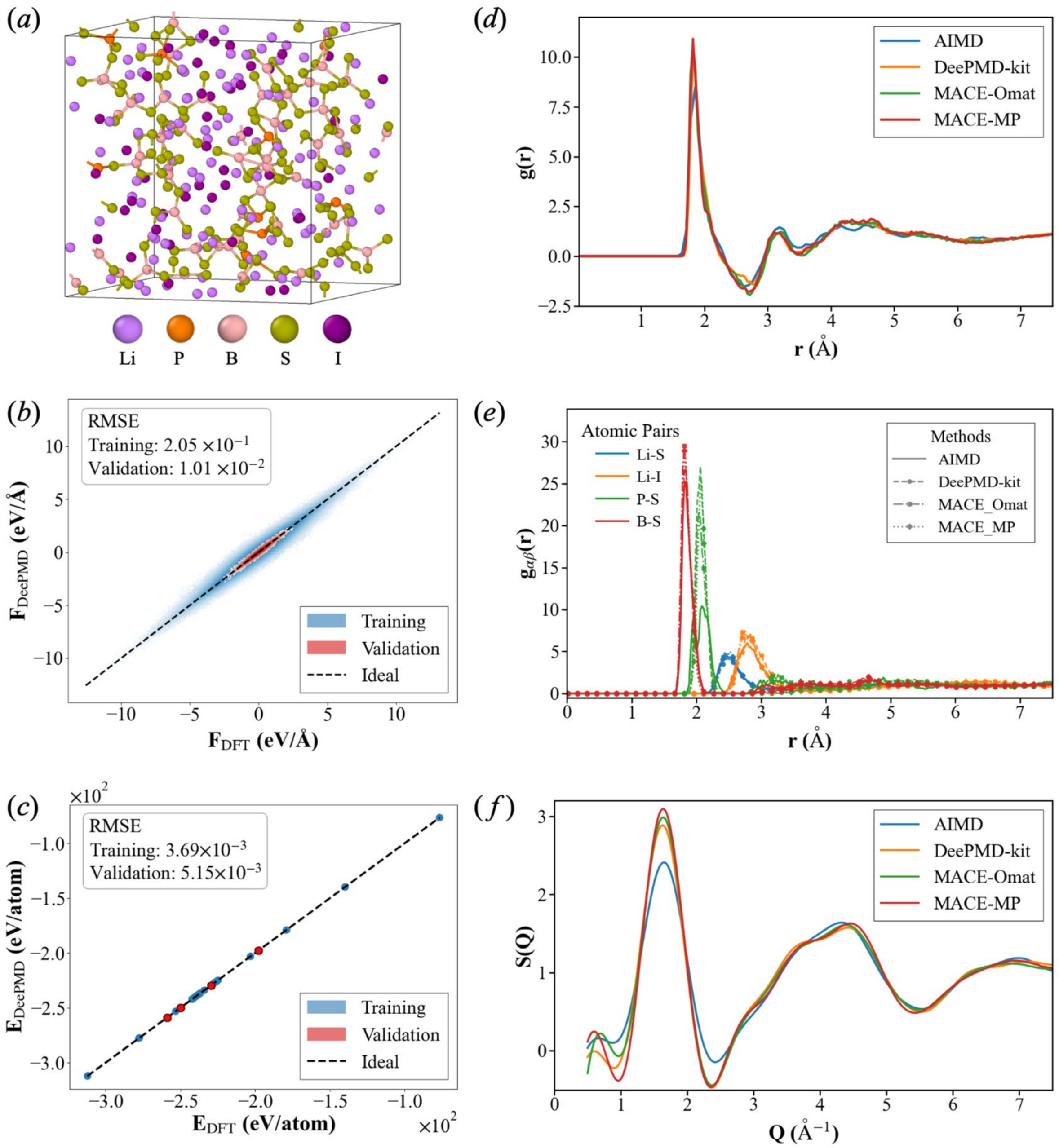

**Figure 1.** Performance evaluation of trained MLIP for LSPBI-*a*5 glassy solid electrolyte. (*a*) Atomic configuration of generated glass structure from AIMD. (*b*-*c*) Parity plots of atomic forces and energies predicted by DeePMD-kit versus DFT reference. (*d*-*f*) Structural properties at 300 K derived from MD simulations using present DeePMD potential, as well as reference MACE-Omat-0 and MACE-MP-0a potentials, benchmarked against AIMD: (*d*) total radial distribution function $g(r)$, (*e*) partial radial distribution functions $g_{\alpha\beta}(r)$ for key atomic pairs, and (*f*) neutron weight structure factor $S(Q)$.

Parity plots of the atomic forces and energies predicted by DeePMD-kit versus DFT reference values are presented in Figures 1*b* and 1*c*, respectively. The MLIP reproduces DFT forces and energies with low root mean square errors (RMSEs) for both training and validation datasets (Figures 1b,c), with the lower validation force error reflecting the smaller forces at 300 K relative to the high-temperature training configurations. To further demonstrate the reliability of our MLIP at elevated temperatures, parity plots for LSPBI-*a*X at 600 K and 1000 K are provided in Supporting Figure S2, which also show excellent agreement. Using this MLIP, we also prepared the same LSPBI-*a*5 glass structure containing 340 atoms,[32] and benchmarked it against two universal MLIPs, namely MACE-MP-0a[33] (primarily trained on crystalline materials) and MACE-Omat-0[34] (trained on extended datasets that involve disordered configurations from high-temperature AIMD trajectories). The simulated densities of LSPBI-*a*5 are 2.13 and 2.21 g/cm$^3$ for MACE-Omat and MACE-MP models, respectively, both slightly lower than the experimental value (2.3 g/cm$^3$). In contrast, the density obtained with the DeePMD model is 2.28 g/cm$^3$, i.e., closer to the experimental value. Importantly, our DeePMD-based model also achieves an approximately eight times higher computational efficiency than the MACE-Omat and MACE-MP models (see Supporting Figure S3). The performance of these general-purpose models, alongside our trained DeePMD-based MLIP, is further evaluated by comparing the the simulated total pair distribution function (Figure 1*d*), partial distribution function (Figure 1*e*) and structure factor (Figure 1*f*). We find that the DeePMD-based model well agrees with the AIMD, MACE-Omat and MACE-MP models and we thus proceed with the DeePMD potential for all subsequent analyses in this work.

The detailed insights into the local bonding of the LSPBI glassy solid electrolyte is gained by using total and partial pair distribution function (PDF) analysis. According to the total PDF in Figure 1*d*, the first sharp peak is located at ~1.85 Å, which is assigned to B-S correlations from trigonal $BS_3$ and tetrahedral $BS_4$. This peak position lies between those of the B-S bond within $BS_3$ trigonal units (~1.81 Å)[35] and that within $BS_4$ tetrahedral units (~1.90 Å).[36] The high intensity of this peak further indicates a well-defined short-range order in the glassy matrix. Notably, a distinct negative peak is observed around 2.75 Å (Figure 1*d*), which arises from the negative neutron scattering length of lithium (-1.90 fm) and suggests that lithium atoms are predominantly located at this characteristic distance from the anionic species (I and S), highlighting the Li-S and Li-I correlations within the first coordination shell. Beyond approximately 5 Å, the PDF oscillates around unity with diminishing amplitude and no long-range periodic fluctuations, confirming the absence of any long-range order in the LSPBI-*a*5 glass samples. Since the content of $P_2S_5$ is significantly lower than that of $B_2S_3$ in the LSPBI-*a*5 composition, its contribution to the total PDF is largely overshadowed, resulting in no distinct peaks corresponding to P-S correlations in Figure 1*d*.

To resolve the individual atomic pair contributions, we present the partial PDFs $g_{\alpha\beta}(r)$ in Figure 1*e* and Supporting Figure S4. The partial PDF for P-S shows a pronounced peak at approximately 2.05 Å, corresponding to the characteristic bond length within $PS_4$ tetrahedral units, confirming the presence of ortho-thiophosphate motifs in the glass structure. Again, the well-defined nature of these peaks indicates a high degree of short-range order around the network-forming cations. Regarding the lithium environment, the Li-S partial PDF exhibits a peak centered at 2.55 Å, while the Li-I correlation appears at a slightly longer distance of 2.85 Å. These values are consistent with typical Li-chalcogen and Li-halogen bond lengths in amorphous sulfide electrolytes.[17] The broader peaks, compared to those of the network formers, reflects the more disordered distribution of mobile $Li^+$ ions within the interstitial sites of the glass matrix. Collectively, the partial PDFs confirm that the local structural motifs obtained from our DeePMD-based model are in excellent agreement with those of the reference AIMD simulations.

Finally, to characterize the medium-range order of the LSPBI glassy electrolytes, we computed the total structure factor $S(q)$ from the MD trajectories, as shown in Figure 1*e* for LSPBI-*a*5. All samples exhibit the characteristic features of non-crystalline materials, including the absence of sharp Bragg peaks and the presence of diffusive scattering patterns. A first intense peak, the so-called first sharp diffraction peak, appears at $q$ of ~1.85 $Å^{-1}$, corresponding to a real space correlation length of ~3.4 Å,[37] i.e., this peak reflects the medium-range ordering in the glass network.

## 2.2 Composition-dependent structural evolution of the glass network

To connect network structure with lithium transport and mechanics, we analyze the evolution of their network topology with varying $P_2S_5$ content. Here, a bridging sulfur is defined as a sulfur atom that connects two network-forming cations (e.g., B-S-B, B-S-P, or P-S-P). A non-bridging sulfur is defined as a sulfur atom bonded to only one network-forming cation (e.g., $B\text{-}S^-$ or $P\text{-}S^-$), typically charge-compensated by $Li^+$. Based on this definition, we analyze the $Q^n$ distribution of $PS_4$ and $BS_4$ tetrahedra (where $n$ = 0-4 denotes the number of bridging sulfur atoms per tetrahedron), as well as the formation of characteristic thiophosphate species across compositions *a*0 to *a*100. Initially, the $PS_4$ tetrahedra evolution is investigated. As shown in Figure 2*a*, the $Q^n$ distribution of $PS_4$ species exhibits a systematic progression with increasing $P_2S_5$ content. At low $P_2S_5$ content, isolated $Q^0$ $PS_4$ units are most abundant, disrupting the B-S framework. With increasing $P_2S_5$, the distribution shifts toward more connected $Q^2$-$Q^4$ $PS_4$ units, with the fraction of $Q^2$ peaking at the *a*25 composition (Figure 2a). Of structural significance is the formation of "meta" thiophosphate configurations, i.e., $[P_2S_6]^{2-}$ dimers, wherein two $PS_4$ tetrahedra interconnect via two shared bridging sulfur atoms. These $[P_2S_6]$ meta units serve as chain-forming nodes that fundamentally alter the network topology, promoting a chain-like or dimer-based architecture at the expense of a fully interconnected three-dimensional network. The fractions of $Q^3$ and $Q^4$ $PS_4$ increase monotonically from 12.9% and 4.0% at *a*5 to 48.3% and 24.3% at *a*100, respectively, indicating a progressive increase in network connectivity. Unlike conventional silicate glasses where high $Q^n$ species imply rigid 3D frameworks, the unique structural chemistry of $[P_2S_6]^{4-}$ units in thiophosphates allows $Q^3$ and $Q^4$ $PS_4$ to participate in flexible, chain-like configurations rather than a stiff 3D network. We further note that, unlike phosphates where $Q^4$ species are experimentally inaccessible due to the P=O double bond, thiophosphates can accommodate $Q^4$ $PS_4$ units because of the lower double-bond character of P=S and the greater polarizability of sulfur. The presence of $Q^4$ $PS_4$ in our simulations is consistent with the known crystal structure of $P_2S_5$ and reflects the distinct network-forming behavior of thiophosphate glasses.

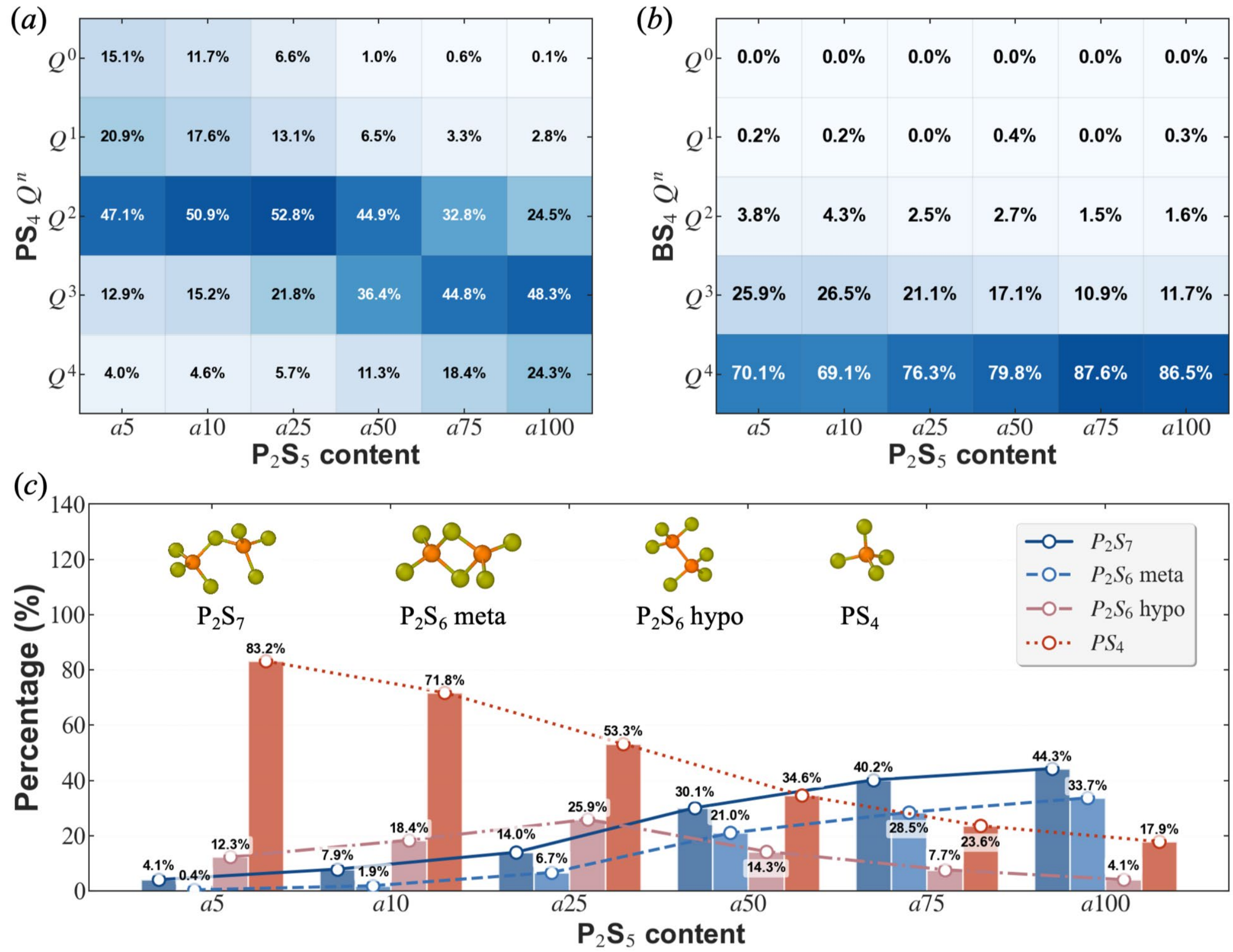

**Figure 2.** Composition-dependent structural evolution in LSPBI glassy electrolyte. (*a*-*b*) $Q^n$ distribution of (*a*) $PS_4$ and (*b*) $BS_4$ units in glassy LSPBI electrolytes with compositions *a*0, *a*5, *a*10, *a*25, *a*50, *a*75, and *a*100. (*c*) Fractions of $P_2S_7$, $P_2S_6$ meta, $P_2S_6$ hypo, and $PS_4$ units in glassy LSPBI electrolytes with compositions *a*0, *a*5, *a*10, *a*25, *a*50, *a*75, and *a*100.

Next, we analyze the compositional evolution of $BS_4$ tetrahedra. In contrast to $PS_4$, the $BS_4$ species remain highly connected across the series (Figure 2*b*). That is, no significant fractions of $Q^0$, $Q^1$, or $Q^2$ $BS_4$ are observed across any composition, consistent with the strong network-forming character of $B_2S_3$. As $P_2S_5$ content increases, the fraction of $Q^3$ $BS_4$ progressively decreases while that of $Q^4$ $BS_4$ correspondingly increases. We note that boron also exists as $BS_3$ units (see Supporting Figure S5 for the $BS_3$/$BS_4$ ratio). Figure 2*c* presents the quantitative distribution of characteristic thiophosphate species ($PS_4$, $P_2S_6$ meta, $P_2S_6$ hypo, and $P_2S_7$) across the compositional series. At low $P_2S_5$ content (*a*5), the network is dominated by $PS_4$ units (83.2% of all P-containing species), consistent with the peak $Q^0$ fraction. We note that $PS_4$ in Figure 2*c* includes all $PS_4$ tetrahedra ($Q^0$-$Q^4$) and only the $Q^0$ fraction represents the truly isolated $[PS_4]^{3-}$ anions, while the remaining $PS_4$ units are connected to the glass network via bridging sulfurs. As the $P_2S_5$ content increases to *a*25, a pronounced structural reorganization occurs, i.e., the $PS_4$ fraction decreases to 53.3%, while $P_2S_6$ meta and $P_2S_7$ begin to emerge. Notably, $P_2S_6$ hypo reaches its maximum fraction of 25.9% at *a*25, exhibiting remarkable correspondence with the $Q^2$ $PS_4$ peak. At higher concentrations (*a*50, *a*75, *a*100), the $PS_4$ fraction declines monotonically to 17.9%, while those of $P_2S_7$ and $P_2S_6$ meta increase to 44.3% and 33.7%, respectively. The dominance of $P_2S_7$ signals the formation of pyro-thiophosphate units and a more condensed network architecture, yet the persistence of $P_2S_6$ meta confirms that the chain-like structural motif remains intact. This compositional evolution agrees with previous experimental Raman spectroscopy data.[16] Finally, we find from the simulated structures that LiI is homogeneously dispersed, without forming large phase-separated aggregates. Iodine atoms are exclusively coordinated by $Li^+$, with no direct I-B or I-P bonding, confirming that LiI acts as an ionic species embedded within the thiophosphate network.

### 2.3 Composition-dependent $Li^+$ hopping

Experimentally, LSPBI-*a*5 xhibits a peak lithium ionic conductivity of 2.4 mS $cm^{-1}$,[16] which has previously[17] been linked to $BS_4$-$PS_4$ corner-sharing in simulations of only the *a*5 composition. We therefore simulate Li transport for *a*0, *a*3, *a*5, and *a*10, as these cover the experimentally observed non-monotonic trend in ionic conductivity, i.e., increasing then decreasing with $P_2S_5$ addition, only manifests at low $P_2S_5$ contents. That is, higher $P_2S_5$ contents (*a*25, *a*50, *a*75, *a*100) lie beyond the experimentally observed conductivity maximum and are therefore excluded from the present diffusion analysis.

We first focus on the long-term (1 ns) variations in lithium-ion displacement magnitudes (Figure 3*a*), with lithium atoms featuring displacement magnitudes exceed 8 Å being highlighted. Fewer lithium ions deviate from their initial positions in LSPBI-*a*10 compared to the baseline LSPBI-*a*0 sample. In contrast, lithium ions in the *a*3 and *a*5 electrolytes exhibit mobility comparable to that observed in *a*0, indicating suppressed Li mobility at higher $P_2S_5$ content. To quantify lithium-ion mobility, the mean square displacement (MSD) of lithium ions is calculated from 300 to 800 K. Figure 3*b* presents the time evolution of MSD for the different glass electrolytes at room temperature. The log-log scale plot indicates that the MSD curves for all samples at room temperature have reached the diffusive regime (see Supporting Figure S6). Interestingly, the addition of a small amount of the glass modifier $P_2S_5$, i.e., in the LSPBI-*a*3 and LSPBI-*a*5 samples, results in a notable increase in the MSD over 1 ns. However, further increasing the $P_2S_5$ content leads to a decrease in lithium mobility, as shown by the LSPBI-*a*10 curve in Figure 3*b*.

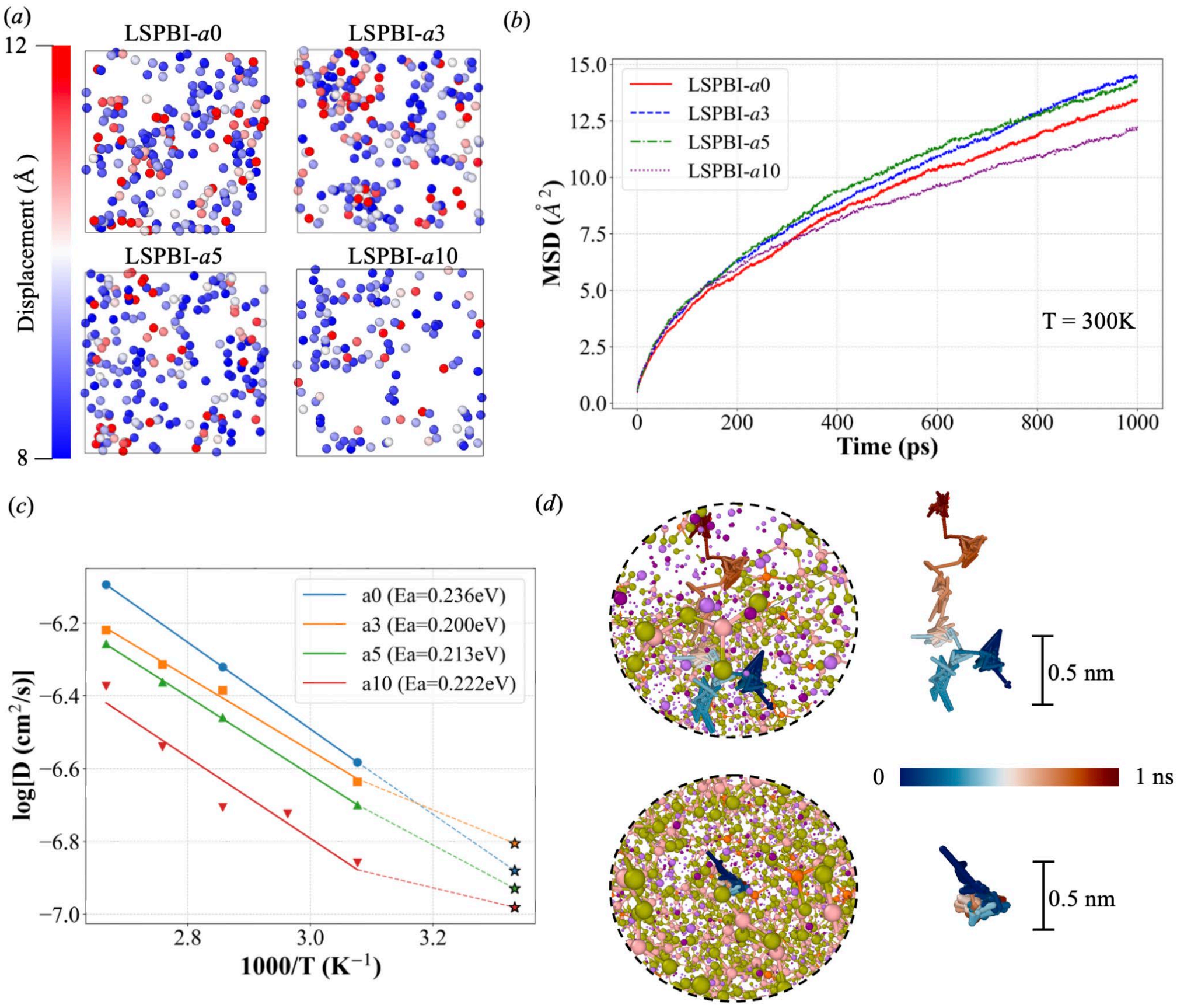

**Figure 3.** Composition-dependent lithium ion hopping in glassy LSPBI electrolyte. (*a*) Displacement magnitudes of lithium ions over 1 ns at room temperature (300 K) in LSBPI *a*0, *a*3, *a*5, and *a*10 samples. Lithium atoms exhibiting displacement magnitudes exceeding 8 Å are highlighted for enhanced visualization. (*b*) MSD of lithium ions in LSBPI *a*0, *a*3, *a*5, and *a*10 samples at 300K. (*c*) Temperature dependent lithium diffusion coefficient. Triangles (325-375 K) are used for Arrhenius fitting; the star (300 K) is the simulation result excluded from the fit. (*d*) Atomic snapshots showing two representative lithium trajectories in the LSPBI-*a*5 glassy solid electrolyte: the upper panel depicts a mobile lithium ion, while the lower panel shows an immobile lithium ion.

To further elucidate the temperature-dependent dynamics, we determine the lithium diffusion coefficients (Figure 3*c*) from the slopes of the MSD vs. time curves in the low-temperature regime (Figure 3*a*). As shown in Supporting Figure S7, all four glassy electrolytes exhibit non-Arrhenius behavior, a characteristic commonly observed in frozen glassy structures as previously reported in various systems.[31,38] The non-Arrhenius curves could be separated into two distinct Arrhenius regimess, with a crossover near 400 K. The high-temperature regime likely involves the activation of additional mobile $Li^+$ ions facilitated by framework relaxation, while the low-temperature regime represents localized hopping of a smaller number of mobile $Li^+$ within the frozen matrix. To quantify ion transport in the glassy state, we focus on the low-temperature regime (from 375 to 325 K) in Figure 3*a*, where the self-diffusion coefficients and activation energies are obtained by linearly fitting the MSD versus time curves. As shown in Figure 3*c*, the activation energy decreases from *a*0 to *a*3 and then increases upon further $P_2S_5$ addition. Although the minimum activation energy in our simulations is observed for the *a*3 rather than the *a*5 composition as reported experimentally,[16] the overall trend is the same. This non-monotonic composition dependence reflects the underlying structural evolution of the glass network: the initial addition of $P_2S_5$ disrupts the continuous B-S framework, creating fragmented domains that facilitate lithium migration and reduce the energy barrier. However, at higher $P_2S_5$ concentrations, the formation of condensed phosphate units such as [$P_2S_6$] and [$P_2S_7$] reconnects the network, reintroducing topological constraints and increasing the activation energy.

The composition-dependent Li mobility follows directly from the network-topology changes established in Section 2.2. In LSPBI-*a*0, $Li^+$ ions migrate through a continuous $BS_3$/$BS_4$-based network, where tortuous pathways and frequent trapping sites limit long-range diffusion. Moderate $P_2S_5$ addition, as in *a*3 and *a*5 compositions, introduces isolated or weakly connected $PS_4$ units that interrupt the B–S framework. This fragmentation reduces topological constraints and creates more percolative $Li^+$ diffusion pathways, as reflected by the extended

Li trajectories in Figure 3*d*, the increased MSDs in Figure 3*b*, and the reduced activation barriers in Figure 3*c*. This mechanism agrees with experimental observations that high conductivity in LSPBI glasses is associated with isolated $PS_4$-rich motifs.[16] Further $P_2S_5$ addition reverses this beneficial effect. In LSPBI-*a*10, the emergence of condensed thiophosphate motifs, including $P_2S_6$ and $P_2S_7$ units, reconnects the network and suppresses long-range $Li^+$ transport. The resulting re-polymerized structure is expected to increase pathway tortuosity and strengthen local $Li^+$ trapping relative to the more fragmented $PS_4$-rich network. Consequently, LSPBI-*a*10 exhibits fewer long-range Li displacements (Figure 3*a*) and a higher activation barrier than the optimally fragmented *a*3/*a*5 compositions (Figure 3*c*).

### 2.4 Brittle to ductile transition

We next examine the mechanical response under uniaxial tensile simulations. As the composition changes from *a*0 to *a*100, a clear transition from brittle to ductile behavior is observed at the nanoscale. Figure 4*a* presents a comparative snapshot of the deformation behavior in two representative LSPBI compositions (*a*5 and *a*100) at various strain levels (0.1, 0.2, 0.3, 0.4, and 0.5). We visualize this through the *Z* component of deformation gradient tensor ($F_Z$) mapping, which quantifies the local stretch along the tensile direction, with $F_Z > 1$ indicating extension and $F_Z < 1$ indicating compression.[39] In LSPBI-*a*5, a pronounced fracture event is observed at higher strain, accompanied by a sharp localization of $F_Z$ near the fracture region, indicative of brittle failure (upper panel of Figure 4*a*, $\varepsilon$=0.5). In contrast, LSPBI-*a*100 exhibits less macroscopic fracture (lower panel of Figure 4*a*). Instead, the formation of some free volume is observed throughout the deformation process and the $F_Z$ distribution remains relatively homogeneous even at large strains, indicating a more nano-ductile response. Figure 4*b* provides an enlarged view of the structural network in LSPBI-*a*100 at a longitudinal strain of 0.5, highlighting the characteristic chain-like structures composed of $BS_3$/$BS_4$ and $PS_4$ units. This morphology suggests that at high $P_2S_5$ content, the glass structural network undergoes a stress-driven atomic reorganization, reconnecting into a chain-like architecture that fundamentally differs from the network topology in low-$P_2S_5$ compositions.

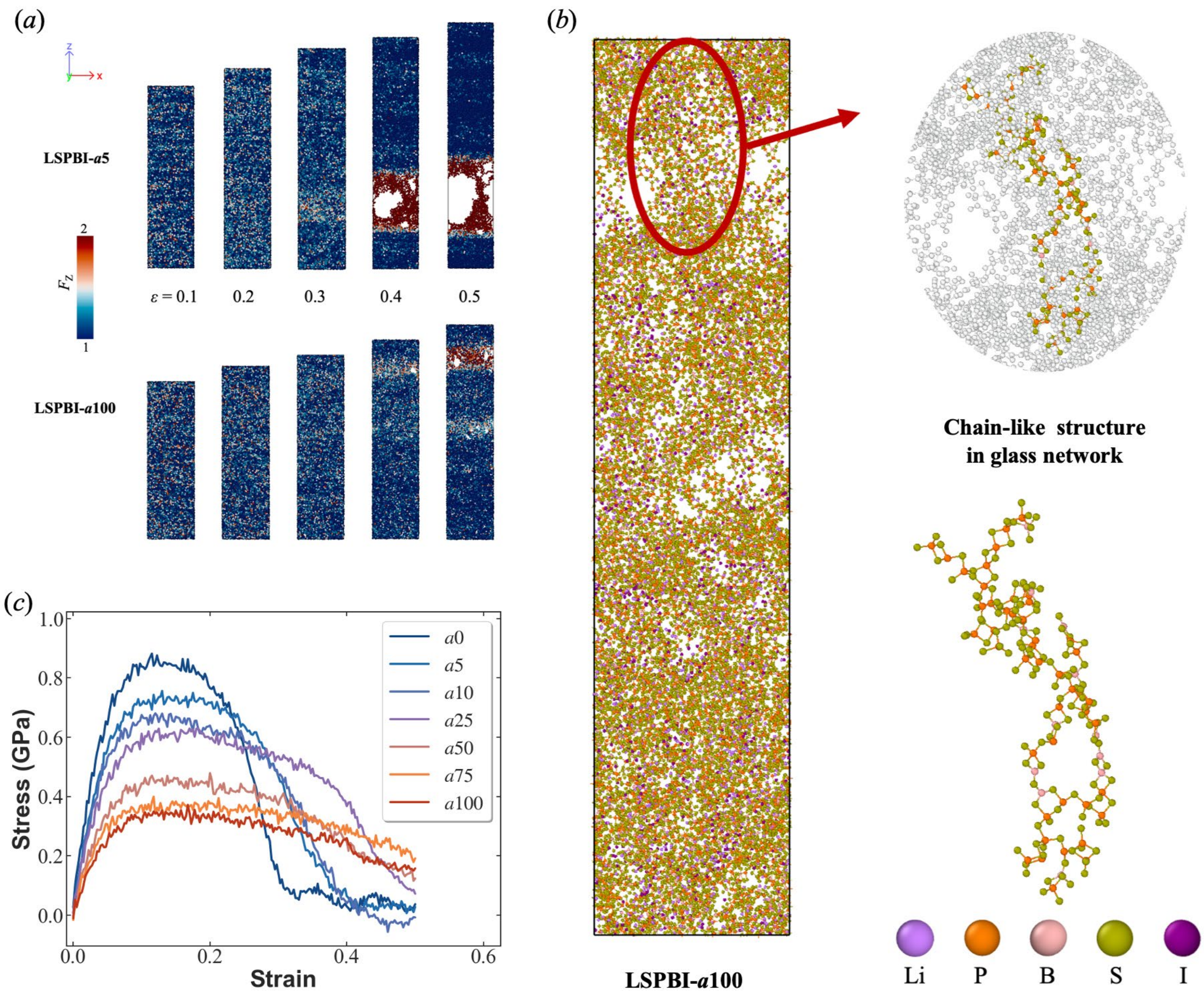

**Figure 4.** Composition-dependent mechanical behavior in glassy LSPBI electrolytes. (*a*) Snapshots of *Z* components of the deformation gradient tensor **F** for two representative LSPBI compositions (*a*5 and *a*100) at strain values of 0.1, 0.2, 0.3, 0.4, and 0.5. (*b*) Atomic snapshot of the glassy LSPBI-*a*100 electrolyte at a strain of 0.5, with an enlarged view highlighting the chain-like structures composed of $BS_3$/$BS_4$ and $PS_4$ units. (*c*) Stress-strain curves of glassy LSPBI electrolytes with compositions *a*0, *a*5, *a*10, *a*25, *a*50, *a*75, and *a*100. The *a*25 composition uniquely combines the high yield strength of *a*10 with the ductility of *a*50.

We then calculate the stress-strain relationships (Figure 4*c*). As the $P_2S_5$ content increases, the maximum peak value of the stress-strain curve, i.e., the tensile strength, exhibits a notable decrease. The replacement of boron by phosphorus introduces weaker P-S bonds (282 to 402 kJ $mol^{-1}$) in place of stronger B-S bonds,[40] which lowers the overall network's resistance to bond stretching and rupture. Generally, higher bond energy results in higher elastic modulus, and thus higher tensile strength.[41] Second, at the level of inelastic deformation under strain, Coulombic interactions between $Li^+$ and the anionic network are likely to be disrupted prior to covalent bond breaking under tensile deformation. These two factors are not mutually exclusive, i.e., the weaker P-S bonds reduce the intrinsic strength of the network, while the disruption of $Li^+$ network Coulombic interactions likely represents the initial inelastic event. The observed decrease in tensile strength with increasing $P_2S_5$ content thus reflects a combination of both effects. In the boron-rich regime, the rigid, highly connected network restricts both $Li^+$ mobility and local structural rearrangement under strain, contributing to its high tensile strength. In contrast, in the phosphorus-rich regime, the more flexible or chain-like network accommodates greater local deformation before bond failure. This mechanical compliance, rather than enhanced $Li^+$ diffusivity, allows stress to be redistributed more effectively and reduces the peak tensile strength. The lower zero-stress $Li^+$ diffusion coefficient in phosphorus-rich compositions (Figure 3*a*) is therefore not contradictory, i.e., under tensile strain, the network itself deforms more readily, passively displacing $Li^+$ rather than requiring active hopping.

More importantly, $P_2S_5$ addition drives a brittle-to-ductile transition. In the B-rich network, topological constraints localize stress and promote brittle fracture,[42] whereas P-containing motifs introduce angular flexibility and energy-dissipating deformation modes. The observed brittle-to-ductile transition can be further understood from the bond angle flexibility, as it is consistent with analogous $Na_2S$-$P_2S_5$-$B_2S_3$ glasses. In their case, broader S-P-S bond angle distributions were linked to softer P-S bonding compared with more rigid B-S units.[40,43] This difference in angular flexibility underpins the distinct mechanical responses observed in our stress-strain simulations, but the incorporation of phosphorus introduces more flexible P-S-P bonding configurations. These bonds have a lower torsional barrier and allow for greater bond angle variation. Upon deformation, this flexible network can dissipate mechanical energy through extensive bond angle bending and bond torsion rather than immediate bond rupture. These microscopic inelastic deformation mechanisms collectively contribute to a larger area under the stress-strain curve (higher toughness) and a more ductile response.

The mechanical transition can be directly linked to the structural motifs identified in Section 2.2. At low $P_2S_5$ content, isolated $PS_4$ units fragment the boron-rich network, but the structure remains dominated by relatively rigid B–S motifs. Stress is therefore transmitted through constrained network regions, promoting strain localization and brittle fracture. With increasing $P_2S_5$ content, $P_2S_6$- and $P_2S_7$-containing motifs introduce flexible P–S–P linkages and chain-like structural units that accommodate tensile strain through bond-angle bending, torsional motion, and local chain rearrangement. These deformation modes distribute strain over a larger volume and suppress catastrophic crack localization. This framework also rationalizes the apparent decoupling between $BS_4$ connectivity and mechanical rigidity. Although the remaining $BS_4$ units become more connected with increasing $P_2S_5$ content, the abundance of B-containing motifs decreases, while P-rich motifs increasingly dominate the deforming network. The brittle-to-ductile transition in Figure 4 therefore reflects a shift in the dominant deformation mechanism, i.e., from stress localization in a constrained B-rich network to distributed strain accommodation in a more flexible, P-rich chain-like network.

## 3. CONCLUSION

This study has uncovered the critical role of network topology in governing the coupled ionic and mechanical properties of multi-anion glass electrolytes for batteries. Using a high-fidelity MLIP validated against AIMD and experimental data, we have systematically explored the ionic transport within the experimentally relevant low-$P_2S_5$ regime (*a*0-*a*10) and the mechanical properties of the LSPBI system across the entire $P_2S_5$ series (*a*0-*a*100). The ionic conductivity exhibits a non-monotonic, composition-dependent trend driven by a structural transition, i.e., initial network fragmentation by isolated [$PS_4$] units enhances $Li^+$ mobility, whereas excessive $P_2S_5$ induces re-polymerization via condensed [$P_2S_6$] and [$P_2S_7$] species, impeding transport. Concurrently, the same structural evolution governs a distinct brittle-to-ductile mechanical transition. The emergence of flexible, chain-like P-S-P architectures overrides the classical correlation between high connectivity and brittleness established for oxide glasses. At the atomic scale, superior damage tolerance in P-rich compositions arises from energy dissipation through bond bending and torsion rather than catastrophic fracture.

Thus, the strategic interplay between network formers can overcome traditional performance trade-offs. Linking dynamical and mechanical characteristics with network topology provides a robust template for investigating other complex glassy electrolytes, such as lithium thiophosphates or halide-based systems. Our work indicates that optimal ion conduction (at *a*5) and maximum ductility (at *a*100) emerge from distinct network topologies. Nevertheless, intermediate compositions (e.g., *a*25-*a*50) offer a pragmatic balance, delivering adequate conductivity while benefiting from improved mechanical robustness.

## 4. METHODS

### 4.1 General methods

Data processing and computational analyses were conducted using established Python scientific libraries, including NumPy[44] for numerical operations, Pandas[45] for data manipulation, and SciPy[46] for advanced scientific computing. Visualization was performed with Matplotlib[47] for generating two-dimensional plots, while atomic-scale structural representations were rendered using Ovito.[48]

### 4.2 *Ab initio* simulations

The initial training dataset for the machine-learned interatomic potential (MLIP) was derived from *ab initio* molecular dynamics (AIMD) simulations. To ensure broad transferability, the dataset encompasses a diverse range of lithium-sulfur-phosphorus-boron-iodine (Li-S-P-B-I) systems, including elemental phases (Li, S, P, B, I), key precursor compounds ($Li_2S$, $P_2S_5$, LiI, $B_2S_3$), and relevant binary/ternary phases ($BI_3$, $Li_2PS_3$, $Li_6PS_5I$, LiB, $P_4S_3I_2$). Furthermore, to represent the compositional space of the target glassy electrolyte, AIMD trajectories for the quaternary system $30Li_2S$-$25B_2S_3$-45LiI-a$P_2S_5$ (with $a$ = 0, 5, 10, 25, 50, 75, and 100) at 3000 K were included. A complete list of all structures is provided in Supporting Table S1. All AIMD simulations were performed within the DFT framework using the CP2K package.[49] The electronic structure was treated with the hybrid Gaussian and plane wave (GPW) method. The PBE functional was employed for exchange-correlation effects, supplemented with Grimme's D3 empirical dispersion correction.[50] Valence electrons were described using the GTH-PBE pseudopotentials with corresponding molecularly optimized basis sets: DZVP-MOLOPT-SR-GTH for Li, B, and I, and TZVP-MOLOPT-GTH for P and S. Convergence of the DFT setup was carefully verified. A plane-wave cutoff of 500 Ry and a relative cutoff of 50 Ry for the Gaussian mapping ensured well-converged total energies. These values are consistent with those shown to achieve high precision in lithium thiophosphate glassy electrolytes.[31] The simulations were carried out in the *NVT* ensemble using a Nosé–Hoover thermostat, with a time step of 0.5 fs. To efficiently sample configurational space and access both crystalline and disordered states, high-temperature AIMD runs were performed at 3000 K for 2.5 ps.

To compare our trained MLIP with AIMD simulations, an LSPBI sample (*a*5) with a systems size of 340 atoms was also prepared using an AIMD-based melt-quench procedure. We adopted a stepwise cooling protocol in which the system was simulated at discrete temperatures: 3000 K, 2500 K, 2000 K, 1500 K, 1000 K, 500 K, and finally 300 K. Each temperature stage was run for 2.5 ps, with the initial configuration for each step taken from the final configuration of the previous temperature using the restart function in CP2K. After the final 300 K stage, we performed an additional *NPT* run followed by an *NVT* run to allow the glassy structure to further relax to its equilibrium state. All other simulation parameters in this melt-quench process were kept identical to those used in the AIMD trajectory simulations as described above.

**4.3 MLIP training and validation process**

The MLIP was developed by using the DeePMD-kit software,[25, 26] with the training and validation dataset calculated from CP2K *ab initio* simulation. The training set included AIMD simulation results at 3000 K for the elements Li, S, P, B, and I, as well as for the primary raw materials of glass electrolytes: $Li_2S$, $P_2S_5$, LiI, and $B_2S_3$. Additionally, to incorporate interatomic interactions, supplementary datasets were added comprising AIMD simulations at 3000 K for BI, $Li_2PS_3$, $Li_6PS_5I$, LiB, $P_4S_3I_2$. Finally, AIMD simulation results for LSPBI at 600 K, 1000 K, and 3000 K were added for various compositions of $30Li_2S$ - $25B_2S_3$ – 45LiI - $aP_2S_5$ ($a$=0, 5, 10, 25, 50, 75, and 100). The validation set was taken as the AIMD simulation results at 300 K for primary raw materials of glass electrolytes: $Li_2S$, $P_2S_5$, LiI, and $B_2S_3$. The full training and validation set can be found in Supporting Table S1. The embedding network and fitting network in the DeePMD-kit were set as three layers with their corresponding neurons as (25, 50, 100) and (240, 240, 240), where two-atom embedding descriptor with a cutoff 6.5 Å was adopted to encode multi-body angular and radial information of neighboring atoms.

MLIP were trained in two sequential stages to optimize model accuracy and transferability. During the first stage, the loss function incorporated energy, force, and virial terms, with their respective prefactors dynamically adjusted from initial values of 0.02, 1000, and 0.02 to final values of 2, 1, and 0.2. The learning rate decayed exponentially from $1\cdot10^{-3}$ to $1\cdot10^{-9}$ over 3,000,000 steps. In the second stage, training was restarted with the loss function again comprising energy, force, and virial terms, whose prefactors evolved from 2, 1, and 0.2 to a uniform value of 1. The learning rate in this stage decreased exponentially from $1\cdot10^{-8}$ to $1\cdot10^{-9}$ across 500,000 steps.

To prevent unphysical atomic overlap at short distances, particularly relevant in high-temperature simulations, the Ziegler-Biersack-Littmark (ZBL) screened nuclear repulsion potential was applied during the first training stage. The ZBL potential is expressed as[51]:

$$V(r)=\frac{1}{4\pi}\frac{Z_1Z_1e^2}{r}\phi\left(\frac{r}{a}\right), \tag{1}$$

with the screening length

$$a=\frac{0.8854a_0}{Z_1^{0.23}+Z_2^{0.23}}, \tag{2}$$

and the screening function

$$\begin{aligned}\phi\left(\frac{r}{a}\right)&=0.1818\exp\left(-3.2\frac{r}{a}\right)+0.5099\exp\left(-0.9432\frac{r}{a}\right)\\&+0.2802\exp\left(-0.4029\frac{r}{a}\right)+0.02817\exp\left(-0.2016\frac{r}{a}\right)\end{aligned}, \tag{3}$$

where $Z_1$ and $Z_2$ represents the atomic numbers of the interacting species, $r$ represents the interatomic distance, $e$ represents the elementary charge, and $a_0$ = 0.529 Å is the Bohr radius. This tabulated ZBL potential was smoothly merged with the deep potential learned interaction via a switching function defined between 0.7 Å and 0.9 Å, using a softmin decay parameter 0.1 to weight the nearest-neighbor distance in the transition. This approach ensures a physically consistent description from the repulsive core to the chemically relevant bonding region. After the first stage, the ZBL potential was removed to allow the model to fully rely on the machine-learned interactions in the second stage.

**4.4 MD simulations of melt-quenched glass**

All molecular dynamics (MD) simulations were performed using Large-scale Atomic/Molecular Massively Parallel Simulator (LAMMPS),[32] with interatomic interactions described by the neural network potential trained via the DeePMD method (as detailed above). Temperature and pressure were controlled using the Nosé–Hoover thermostat and barostat, respectively. To obtain the glassy LSPBI solid electrolytes, samples with compositions of $30Li_2S$-$25B_2S_3$-45LiI-$aP_2S_5$ ($a$ = 0, 5, 10, 25, 50, 75, 100) were first constructed by randomly packing $Li_2S$, $B_2S_3$, LiI, and $P_2S_5$ units into a periodic simulation cells by using PACKMOL package[52]. These initial configurations were then subjected to a melt-quench procedure. That is, each system was heated to 1200 K, held for 100 ps, and subsequently cooled to 300 K at a uniform rate of 2.5 K/ps under the *NPT* ensemble, followed by an equilibration period in the *NPT* ensemble for 50 ps and another equilibration period in the *NVT* ensemble for 50 ps. A time step of 0.5 fs was used throughout all stages to ensure numerical stability. Although the simulation cell dimensions differ between the lithium-diffusion and mechanical-property studies, the same melt-quench protocol was applied consistently. The detailed composition (number of each molecule) and simulation-box dimensions for every system are provided in Supporting Tables S2 and S3.

**4.5 Identification and counting of local structural units**

To quantify the connectivity of phosphorus-containing units in the LSPBI glass network, $PS_4$ tetrahedra were identified based on P-S coordination distance. Bridging sulfur atoms were defined as those bonded to two network-forming cations (P or B). The $Q^n$ distribution counts every $PS_4$ tetrahedron according to the number of bridging sulfur atoms it possesses, regardless of whether the bridging partner is another $PS_4$ or a $BS_4$/$BS_3$ unit. For the speciation analysis distinguishing isolated $PS_4$ units from $P_2S_7$ dimers (two $PS_4$ sharing one bridging sulfur), a weight-based counting scheme was adopted to avoid double-counting when a $PS_4$ unit connects to multiple other $PS_4$ units. In this scheme, each phosphorus atom participating in *n* P-S-P bridges contributes a weight of 1/*n* to each connected dimer. The fraction of each species was then calculated as the summed weighted contributions over all phosphorus atoms in the system.

**4.6 Structural descriptors**

The simulated LSPBI glass electrolyte structures were analyzed by computing the partial pair distribution function (PDF) $g_{\alpha\beta}(r)$ between the two types of atoms $\alpha$ and $\beta$, with the expression described as, [53]

$$g_{\alpha\beta}(r) = \frac{1}{\rho_\beta} \cdot \frac{dn_{\alpha\beta}(r)}{4\pi r^2 dr}, \tag{4}$$

where $dn_{\alpha\beta}$(r) represents the number of $\beta$ atoms in spherical shell between $r$ and $r+dr$ around $\alpha$ atoms, $\rho_\beta$ represents the number density of $\beta$ atoms. Then the total PDF $g(r)$ was expressed as a function of partial PDF $g_{\alpha\beta}(r)$, [53]

$$g(r) = \frac{\sum_\alpha \sum_\beta c_\alpha c_\beta b_\alpha b_\beta g_{\alpha\beta}(r)}{\left(\sum_\alpha c_\alpha b_\alpha\right)^2}, \tag{5}$$

where $c_\alpha$ and $c_\beta$ represents the atomic fraction of element $\alpha$ and $\beta$ in the system, $b_\alpha$ and $b_\beta$ represents the amplitude of neutron scattering by the atomic nucleus of element $\alpha$ and $\beta$, with their values equal to -1.90 fm, 2.847 fm, 5.13 fm, 5.30 fm, 5.28 fm for Li, S, P, B, I respectively.

Another structural descriptor used in this paper is the neutron weight structure factor $S(q)$, which was calculated by using the total pair distribution function,[54]

$$S(q) = 1 + 4\pi\rho \int_0^{r_{\max}} r^2 \frac{\sin(qr)}{qr} \left(g(r) - 1\right) dr, \tag{6}$$

where $q$ represents the scattering vector magnitude, $r_{\max}$ represents the maximum distance of integration and typically is set as the half of the box size, and $\rho$ represents the atoms number density.

### 4.7 Self-lithium diffusion

To quantitatively evaluate lithium-ion transport, we calculated the self-diffusion coefficients of $Li^+$ in LSPBI electrolytes with compositions of $a0$, $a3$, $a5$, and $a10$. Molecular dynamics simulations were performed in a cubic cell; detailed information regarding the simulation box dimensions and atom counts is provided in Supporting Table S2. The lithium ionic conductivity of the LSPBI glassy electrolytes was calculated by using the mean square displacement (MSD). The as-prepared glass samples LSPBI $a0$, $a3$, $a5$, $a10$ were equilibrated in the *NVT* ensemble at different temperature for 1 ns to generate trajectories, which was then used to calculate MSD as,

$$\mathrm{MSD}(t) = \left\langle \left| \vec{r}_i(t) - \vec{r}_i(0) \right|^2 \right\rangle, \tag{7}$$

where $\vec{r}_i(t)$ represents the position vector of the $i^{\text{th}}$ atom at time t, and $\vec{r}_i(0)$ represents the position vector of the $i^{\text{th}}$ atom initially. For a three dimensional glassy sample, the self-diffusion coefficient of Li atoms can be derived by calculating the slope of MSD as,

$$D = \frac{1}{6} \lim_{t \to \infty} \frac{\mathrm{MSD}(t)}{t}. \tag{8}$$

The activation energy $E_a$ was calculated by fitting the Arrhenius function as,

$$D(T) = D_0 \exp\left( \frac{-E_a}{k_B T} \right), \tag{9}$$

where $k_B$ represents the Boltzmann constant, and T represents the temperature, and $D_0$ represents the self-diffusion coefficient at an infinite temperature.

The MSD of lithium ions was monitored over a wide temperature range (from 800 K down to 325 K) for 1 ns to capture both high-temperature and low-temperature dynamics. As illustrated in Supporting Figure S7, all samples exhibit non-Arrhenius behavior with a dynamic transition around 400 K, a characteristic feature of frozen glassy structures. Accordingly, the extraction of diffusion coefficients in the low-temperature regime was carefully optimized on a per-composition basis. For samples such as LSPBI-$a0$, which display a well-defined linear Arrhenius region at low temperatures, three representative temperatures (375 K, 350 K, and 325 K) were sufficient to reliably extract the diffusion activation energy. For compositions where the Arrhenius plot exhibits more pronounced curvature near the glass transition region (around 400 K), we refined the temperature grid by including additional points (362.5 K and 337.5 K). This refinement ensured that the linear fits were confined exclusively to the truly "frozen" glassy state, excluding the high temperature regime where structural relaxation set in and leads to a higher apparent activation energy due to the activation of additional mobile $Li^+$ ions or cooperative migration mechanisms. Finally, the lithium self-diffusion coefficients were obtained by linearly fitting the MSD curves within this carefully defined low-temperature regime, where diffusive behavior is well-established and minimally perturbed by thermal fluctuations. This approach guaranteed that the calculated activation energies and pre-exponential factors accurately reflect the long-range ion hopping barriers in the rigid glassy matrix, rather than being convoluted with temperature-dependent structural changes.

### 4.8 Tensile fracture simulations

To investigate the mechanical response, uniaxial tensile deformation was simulated for LSPBI electrolytes with varying $P_2S_5$ content, specifically compositions *a*0, *a*5, *a*10, *a*25, *a*50, *a*75, and *a*100. Given that subtle variations in the glass network former ratio may have negligible impact on the tensile behavior, the intermediate *a*3 composition was excluded from the mechanical analysis. Detailed information regarding the simulation box dimensions and atom counts is provided in Supporting Table S3 Prior to loading, the simulation cells were constructed with an elongated geometry along the tensile direction (*z*-axis) using Packmol, following the same procedure described in Section 4.4 for bulk glass generation. The as-prepared cells were then equilibrated in the *NPT* ensemble at 300 K and zero pressure for 100 ps before applying tensile deformation. To establish an appropriate loading rate, a systematic strain rate sensitivity study was conducted across two orders of magnitude: $5\cdot10^8$ $s^{-1}$, $1\cdot10^9$ $s^{-1}$, $5\cdot10^9$ $s^{-1}$, $1\cdot10^{10}$ $s^{-1}$, and $5\cdot10^{10}$ $s^{-1}$. As summarized in Supporting Figure S8, while the magnitude of the stress response exhibits some rate dependence, the relative ranking of mechanical performance among different compositions remains consistent across all tested strain rates, validating the comparative analysis. Additionally, finite-size effects were evaluated by simulating the *a*5 composition with system sizes ranging from 5,000 to 50,000 atoms (Supporting Figure S9). The stress-strain curves show negligible variation beyond 20,000 atoms, confirming that the selected model size of 20,000 atoms is sufficient to capture representative mechanical behavior.

Based on these validations, a strain rate of $5 \cdot 10^9$ $s^{-1}$ was adopted for all subsequent tensile simulations. During loading, uniaxial tension was applied along the *z*-direction by continuously rescaling the box dimensions, while the lateral dimensions (*x* and *y*) were held fixed. This loading configuration corresponds to a uniaxial strain condition, as the Poisson contraction in the transverse directions is constrained by the fixed lateral box dimensions. The final stress-strain curve is obtained by averaging over three MD simulations. The system was maintained at 300 K using an *NVT* ensemble throughout the deformation process. Atomic trajectories were recorded at strain intervals of 0.001, and the engineering stress along the loading direction was extracted to construct the stress-strain curves for further analysis of composition-dependent mechanical properties. Supporting Figure S10 systematically presents the simulated mechanical response of compositions LSPBI-*a*0 through LSPBI-*a*100 under applied strains ranging from 0.1 to 0.5.

**ACKNOWLEDGEMENTS**

This work was supported by a MSCA Postdoctoral Fellowship (101148843) from Horizon Europe. T.J. and K.T. acknowledge support from the Novo Nordisk Foundation (NNF23OC0087524).

**DATA AND CODE AVAILABILITY**

The simulation workflow templates for generating LSPBI glass structures and executing molecular dynamics runs, along with the trained neural network potential, are publicly accessible on GitHub at https://github.com/clyongAAU/AAU_glass_LSPBI. All relevant source data supporting the findings of this study are included in the paper and its supplementary information. The simulations in this work were performed using LAMMPS, a widely recognized molecular dynamics package, and the DeePMD-kit for machine-learning interatomic potentials. Both tools are publicly accessible under open-source licenses, with LAMMPS available at https://www.lammps.org and DeePMD-kit hosted at https://github.com/deepmodeling/deepmd-kit.

# Supporting Information

*for*

## Local Structure Dictates Ionic Transport and Mechanical Properties in Glassy Solid Electrolytes for Lithium Batteries

Yong Li [a], Tao Du [b], Rasmus Christensen [a,c], Timothée Jamin [a], Zhencai Li [a], Qi Zhang [d], Xiaoyi Xu [a], Kasper Tolborg [a], Yuanzheng Yue [a], Morten M. Smedskjaer [a,*]

[a] *Department of Chemistry and Bioscience, Aalborg University, 9220 Aalborg East, Denmark*

[b] *Department of Applied Physics, The Hong Kong Polytechnic University, Kowloon, Hong Kong 999077, China*

[c] *Department of Applied Physics, Tohoku University, Sendai, Japan*

[d] *Corning Research Center in China, Corning Incorporated, Shanghai, China*

[*] *Corresponding author. E-mail: mos@bio.aau.dk*

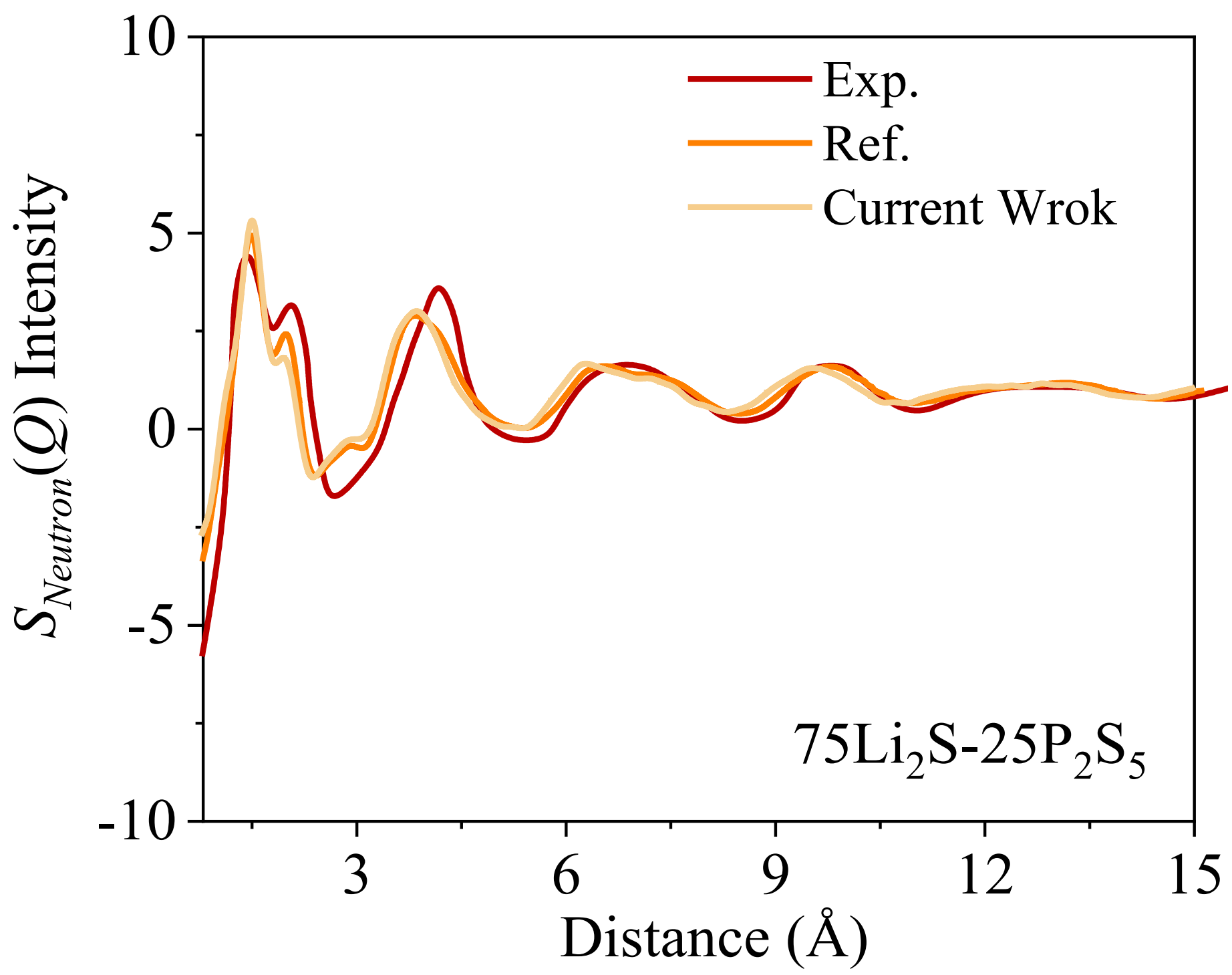

**Figure S1.** Comparison of the neutron weighted structure factor $S_{Neutron}(Q)$ in $75Li_2S$ - $25P_2S_5$ from this study with results from existing experimental and simulation studies.[1, 2]

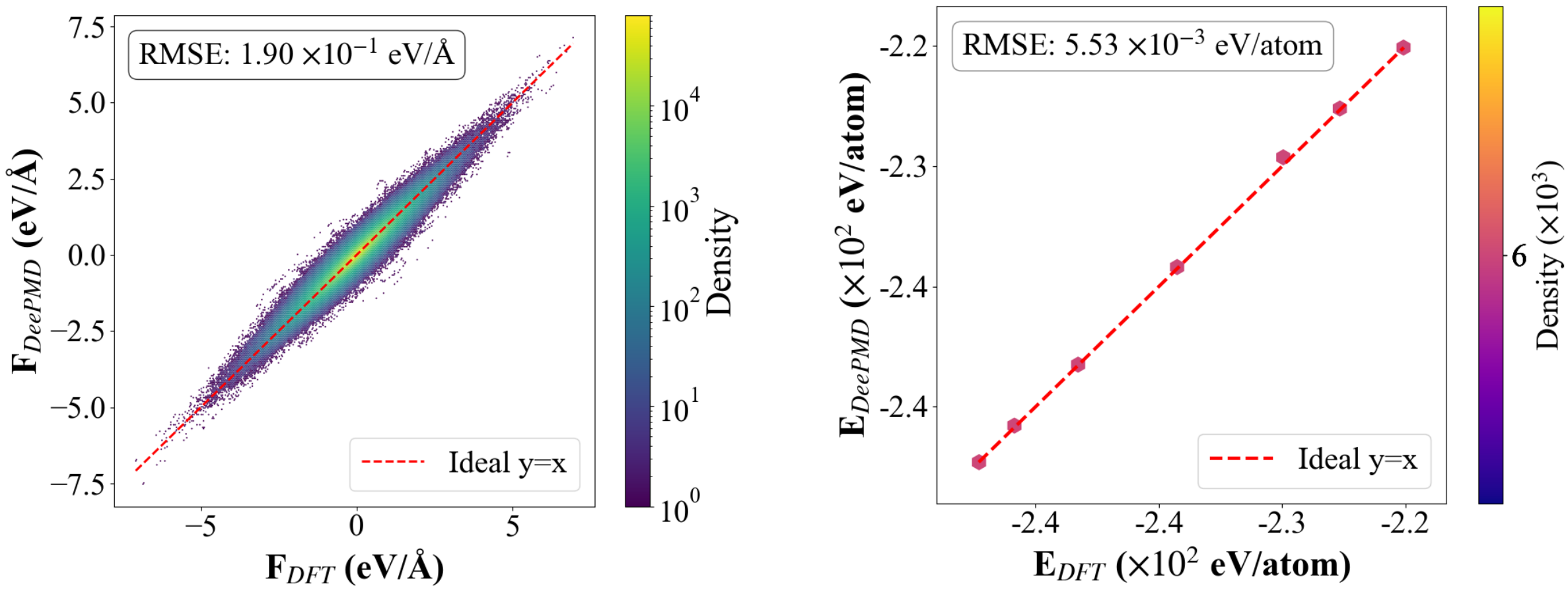

**Figure S2.** Parity plots comparing atomic forces (left panel) and energies (right panel) predicted by the DeePMD-kit MLIP with DFT reference values for LSPBI-*a*0, *a*5, *a*10, *a*25, *a*50, *a*75, and *a*100 at 600 K and 1000 K. It should be noted that the DFT reference data used here are independent of the training dataset for the same compositions and temperatures, i.e., they are derived from separate AIMD trajectories not seen during training.

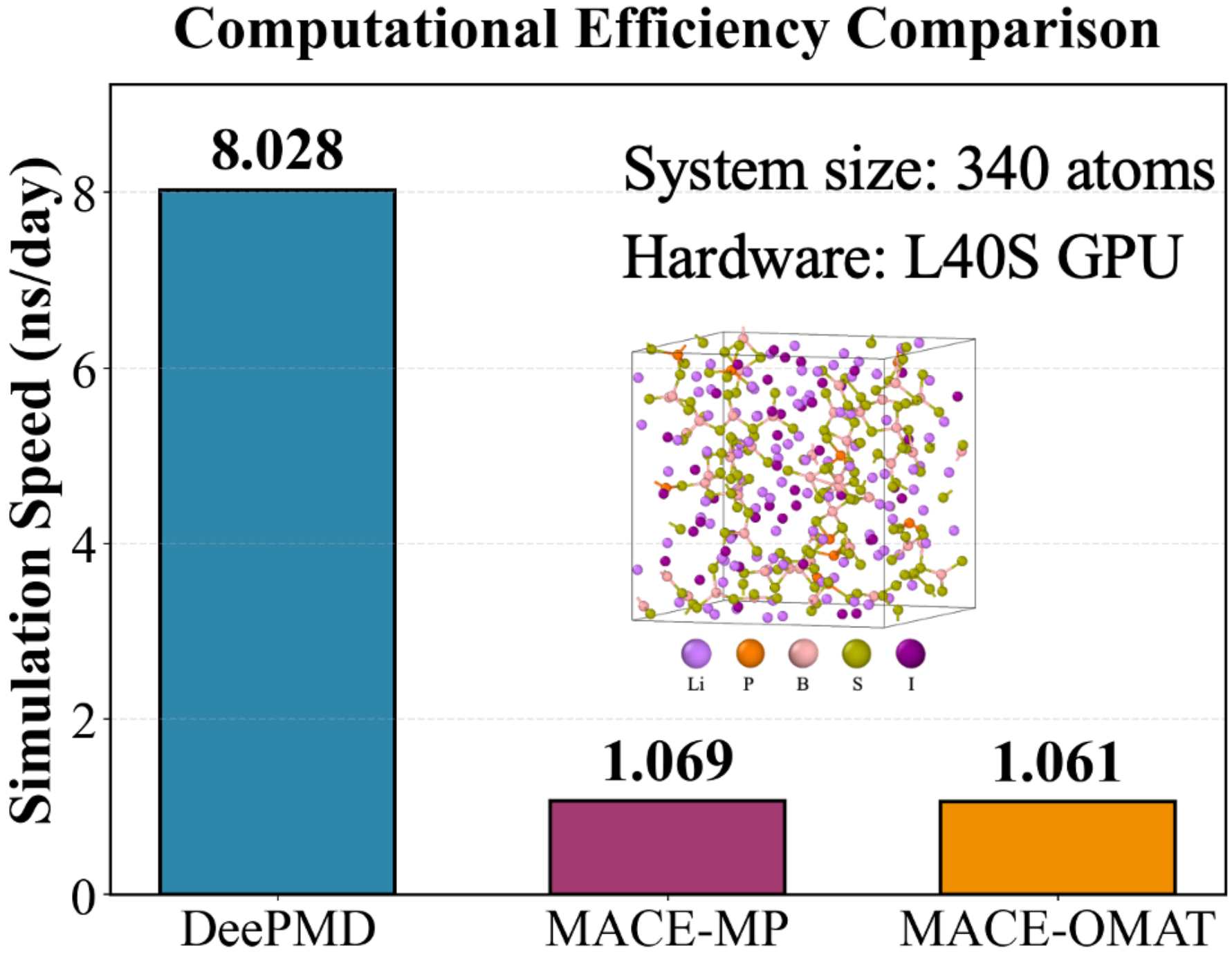

**Figure S3.** Computational efficiency comparison of DeePMD, MACE-MP-0a, and MACE-OMAT-0 machine learning interatomic potentials for simulating LSPBI glassy electrolytes.

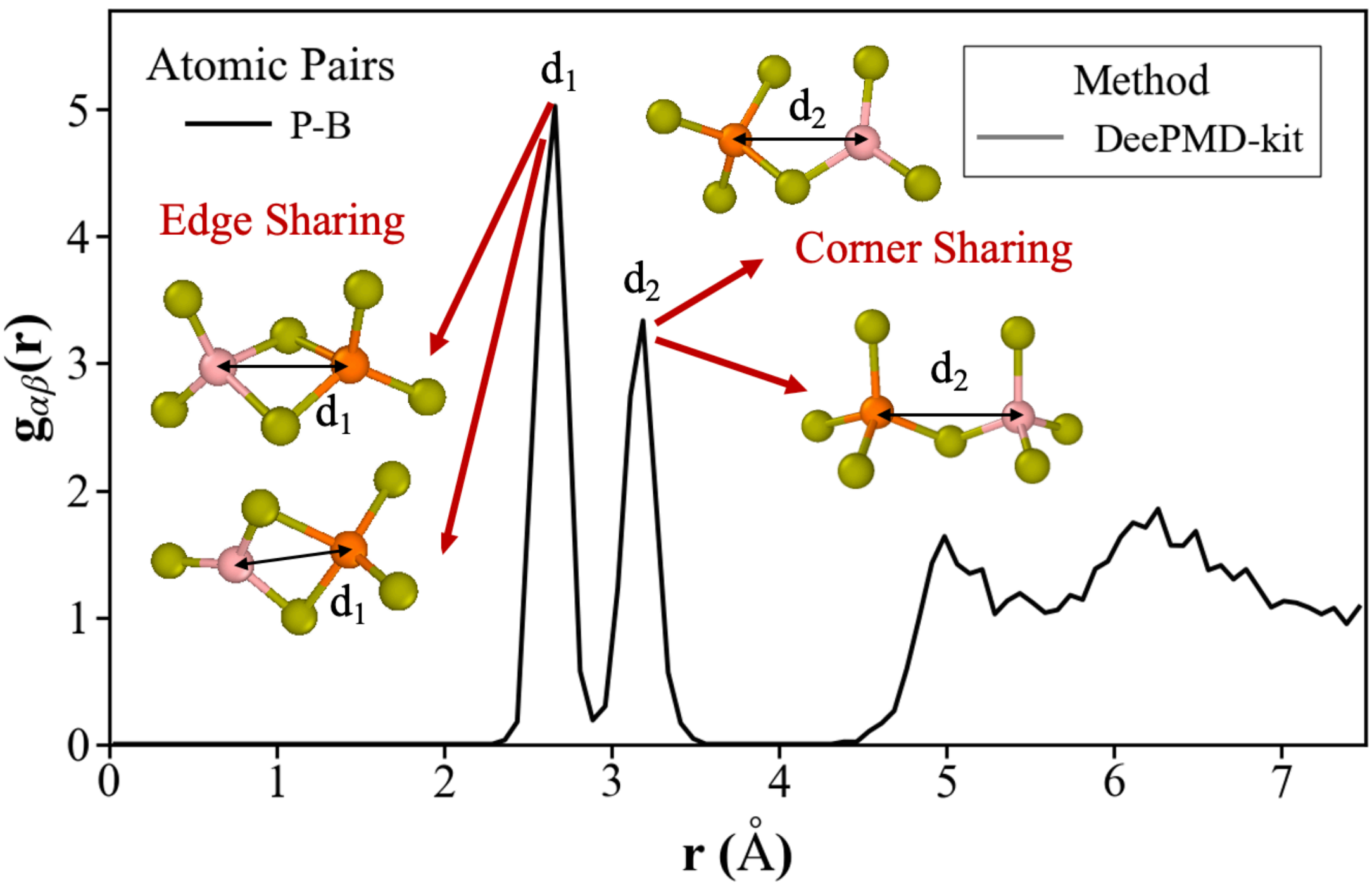

**Figure S4.** Partial pair distribution function for B-P calculated from the trained DeePMD-based MLIP. The first and second peaks correspond to B-P correlations in tetrahedral configurations linked via edge-sharing and corner-sharing motifs, respectively.

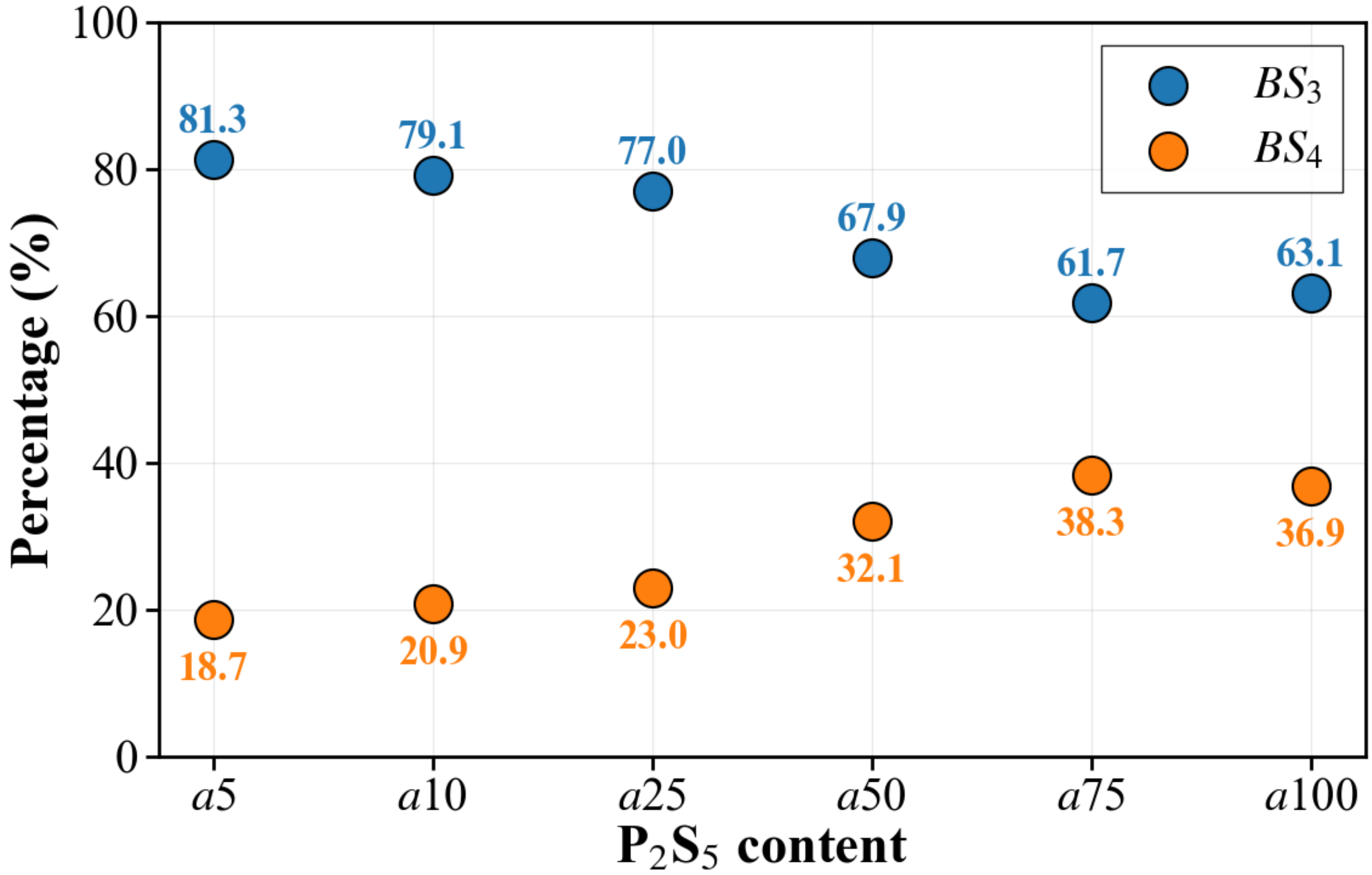

**Figure S5.** Composition dependence of the $BS_3$ and $BS_4$ ratios in glassy LSPBI electrolytes (from *a*5 to *a*100).

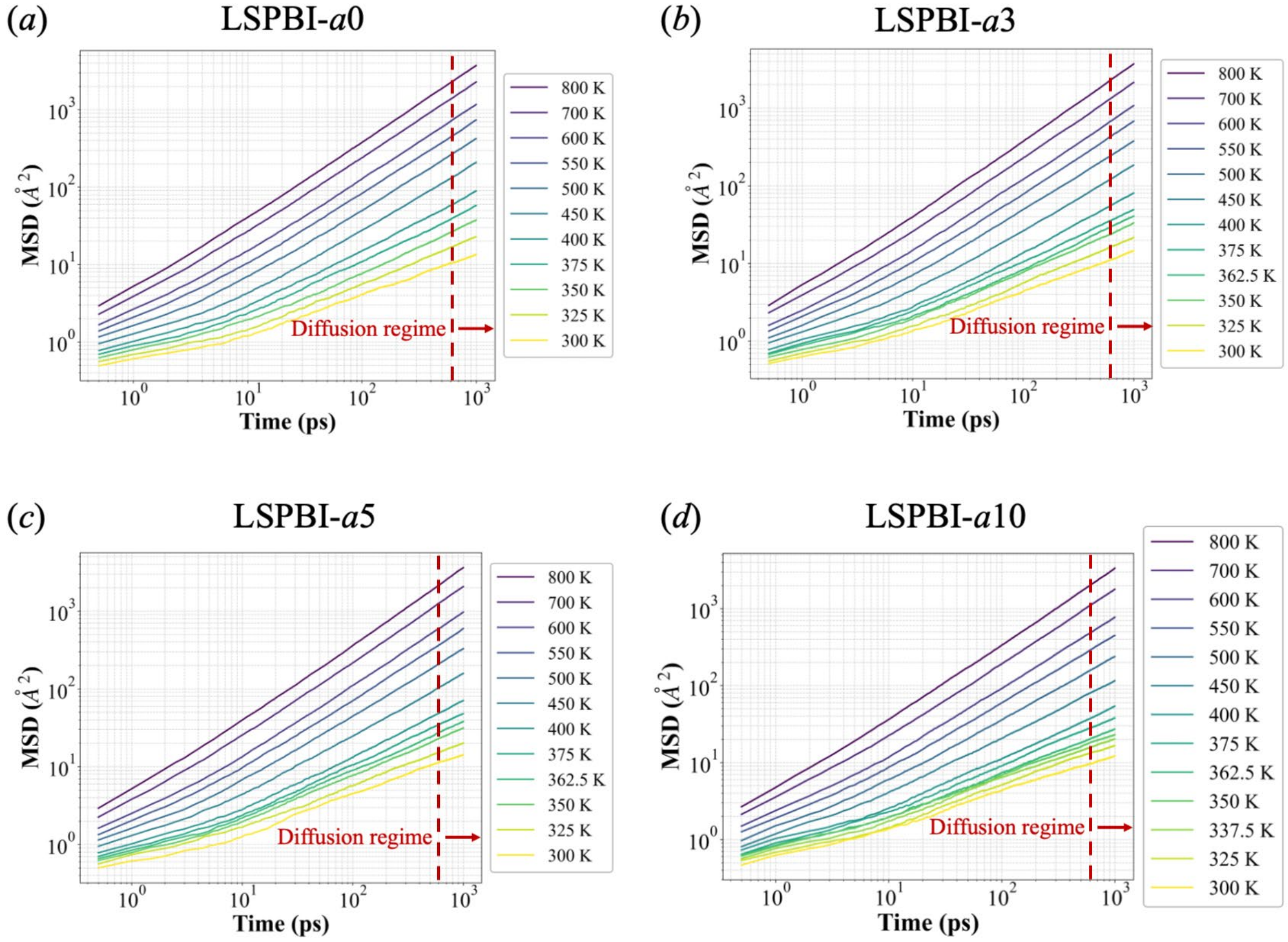

**Figure S6.** Mean squared displacement (MSD) of various solid glassy electrolytes presented on a log-log scale: (*a*) LSPBI-*a*0, (*b*) LSPBI-*a*3, (*c*) LSPBI-*a*5, and (*d*) LSPBI-*a*10.

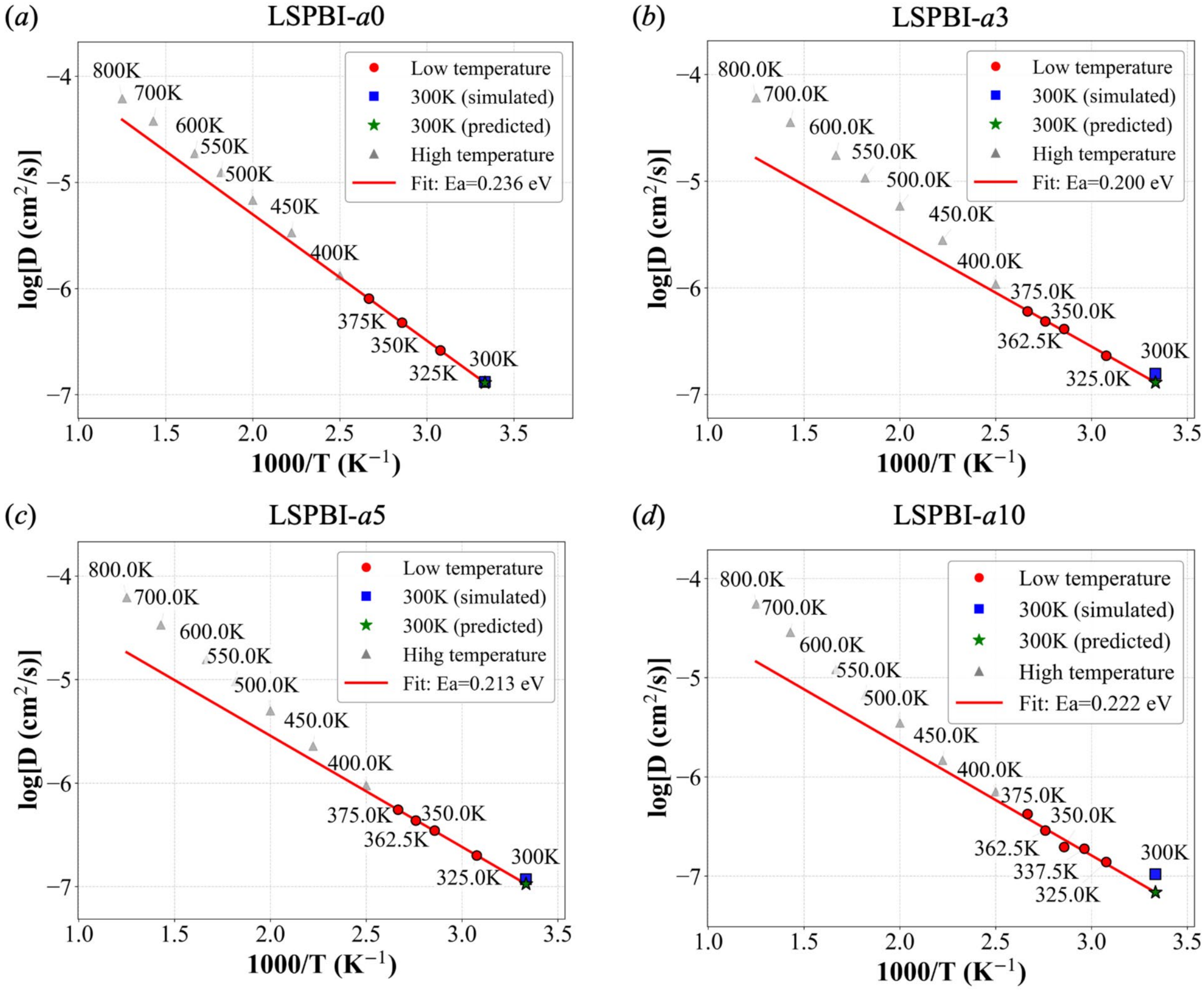

**Figure S7.** Self-diffusion coefficients of various solid glassy electrolytes: (*a*) LSPBI-*a*0, (*b*) LSPBI-*a*3, (*c*) LSPBI-*a*5, and (*d*) LSPBI-*a*10. Diffusion behavior was examined over a wide temperature range, including high temperatures (800 K, 700 K, 600 K, 550 K, 500 K, 450 K, 400 K) and low temperatures (375 K, 362.5 K, 350 K, 337.5 K, 325 K). The lithium self-diffusion coefficients were linearly fitted using the data from the relatively low-temperature range.

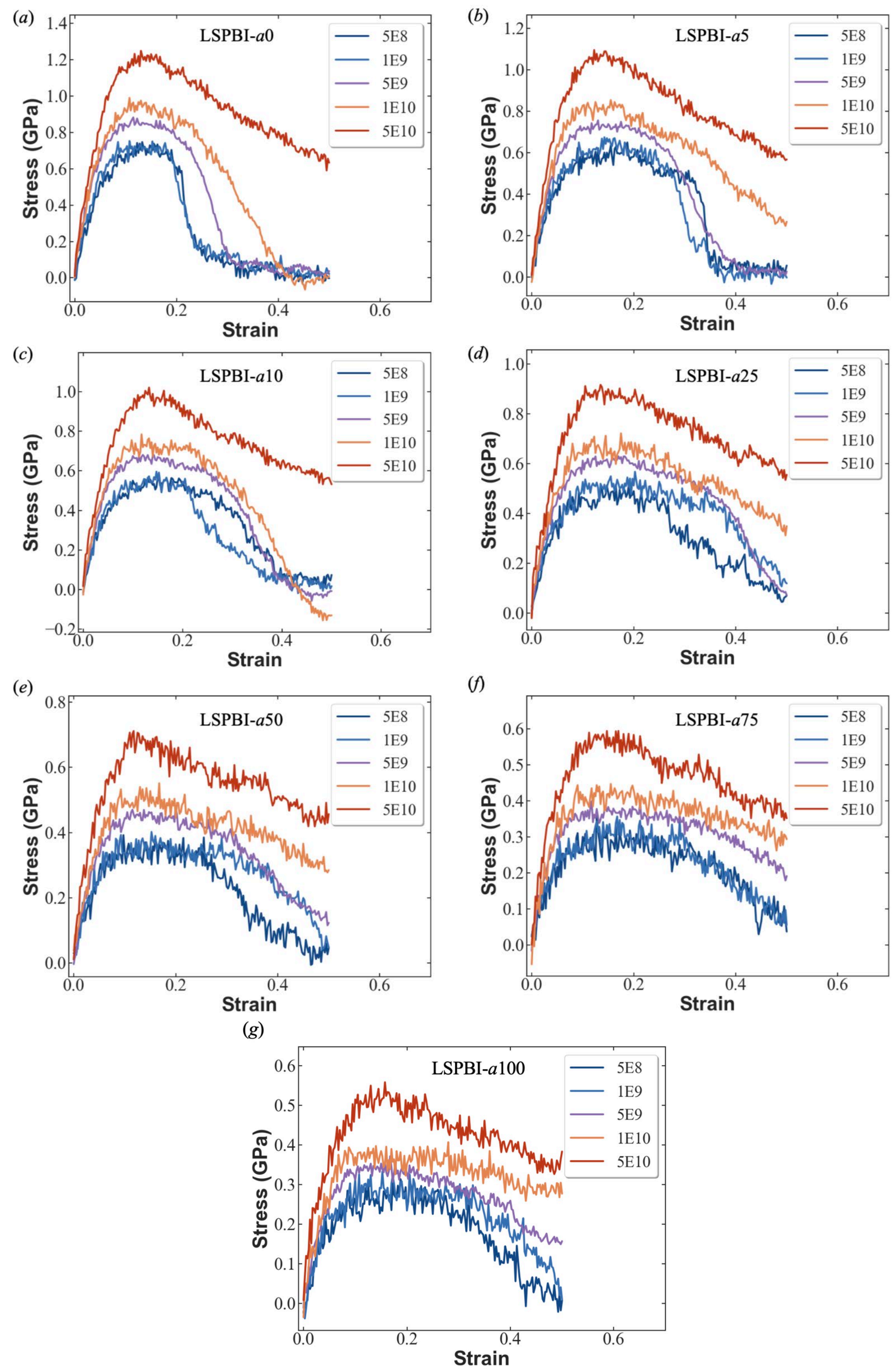

**Figure S8.** Effects of strain rate on the simulated stress-strain curve of various solid glassy electrolytes: (*a*) LSPBI-*a*0, (*b*) LSPBI-*a*5, (*c*) LSPBI-*a*10, (*d*) LSPBI-*a*25, (*e*) LSPBI-*a*25, (*f*) LSPBI-*a*50, and (*g*) LSPBI-*a*100.

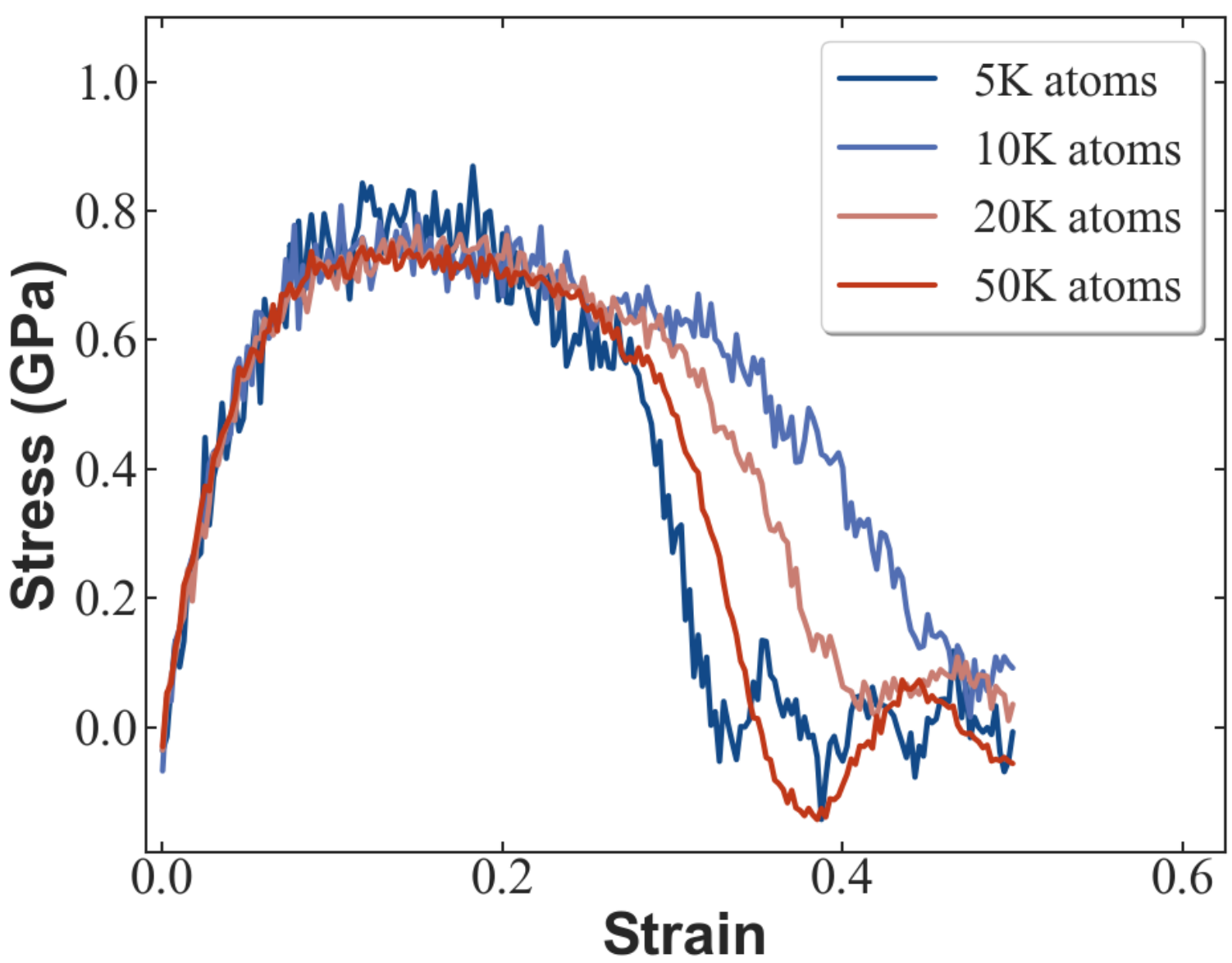

**Figure S9.** Effect of system size on the simulated stress-strain curve of the LSPBI-*a*5 solid glassy electrolyte. Results are shown for models containing 5,000, 10,000, 20,000, and 50,000 atoms.

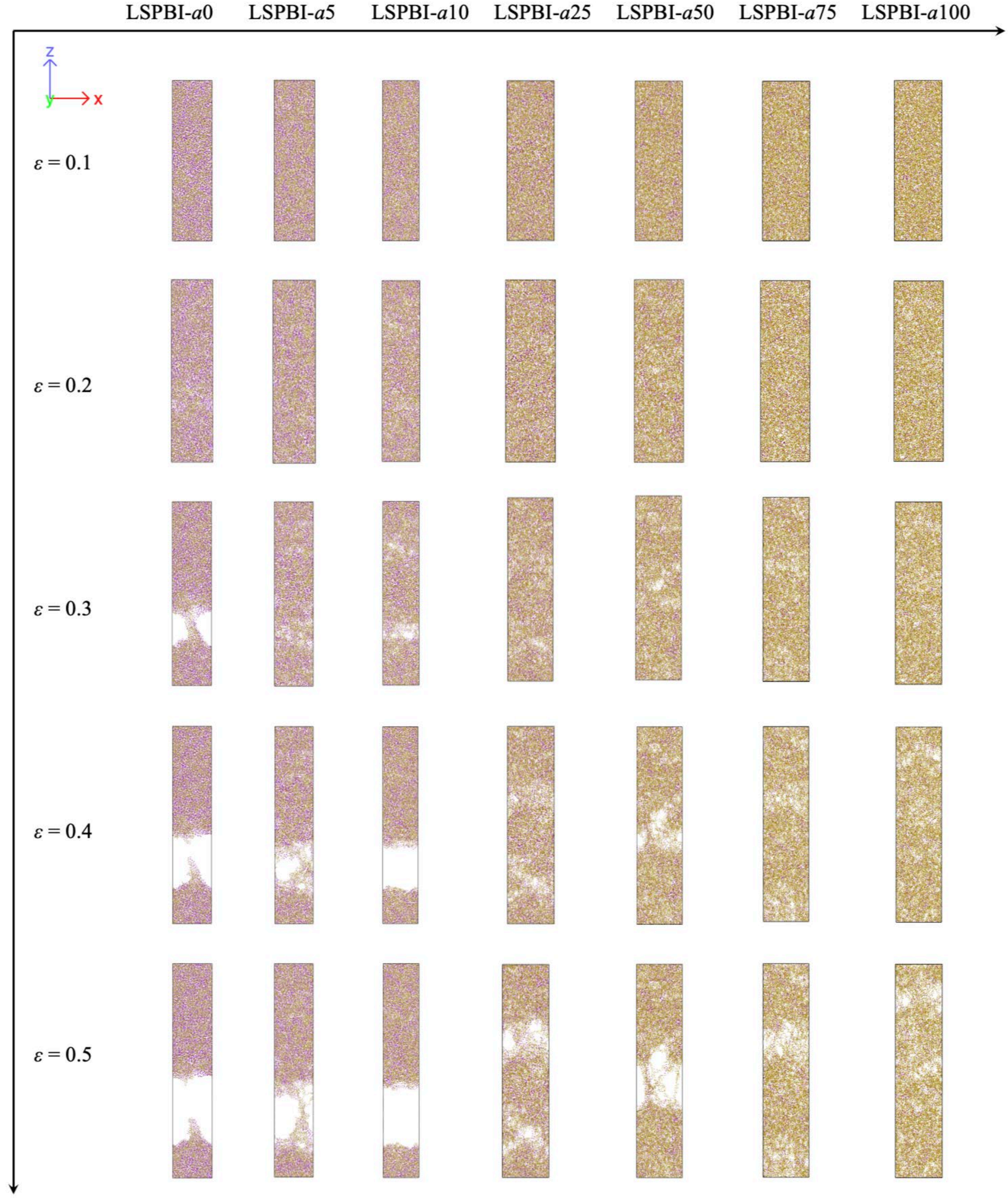

**Figure S10.** Atomic snapshots capturing the fracture process of various LSPBI solid glassy electrolytes (*a*0, *a*5, *a*10, *a*25, *a*50, *a*75, *a*100) at increasing strain levels of 0.1, 0.2, 0.3, 0.4, and 0.5.

**Table S1.** Overview of training and validation dataset for training the machine learning interatomic potential.

***Training dataset:***

| Composition | System | Simulation Temperature | Snapshot of initial simulation structure |
| --- | --- | --- | --- |
| Li | Cubic | 3000K | |
| S | Monoclinic | 3000K | |
| B | Cubic | 3000K | |
| I | Tetragonal | 3000K | |
| P | Monoclinic | 3000K | |
| $B_2S_3$ | Trigonal | 3000K | |
| $B_2S_3$ | Monoclinic | 3000K | |
| $B_2S_3$ | Tetragonal | 3000K | |
| $Li_2S$ | Cubic | 3000K | |
| LiI | Cubic | 3000K | |
| $P_2S_5$ | Triclinic | 3000K | |

| $BI_3$ | Hexagonal | 3000K | |
|---|---|---|---|
| LiB | Cubic | 3000K | |
| $Li_2PS_3$ | Hexagonal | 3000K | |
| $Li_2PS_3$ | Orthorhombic | 3000K | |
| $Li_6PS_5I$ | Cubic | 3000K | |
| $Li_6PS_5I$ | Monoclinic | 3000K | |
| $P_4S_3I_2$ | Orthorhombic | 3000K | |
| $P_4S_3I_2$ | Triclinic | 3000K | |
| LSPBI_a0 | Amorphous | 300K, 600K, 1000K, 3000K | |
| LSPBI_a5 | Amorphous | 300K, 600K, 1000K, 3000K | |
| LSPBI_a10 | Amorphous | 300K, 600K, 1000K, 3000K | |
| LSPBI_a25 | Amorphous | 300K, 600K, 1000K, 3000K | |
| LSPBI_a50 | Amorphous | 300K, 600K, 1000K, 3000K | |
| LSPBI_a75 | Amorphous | 300K, 600K, 1000K, 3000K | |

| LSPBI_a100 | Amorphous | 300K, 600K, 1000K, 3000K | |
|---|---|---|---|

***Validation dataset:***

| $B_2S_3$ | Trigonal | 300K | |
|---|---|---|---|
| $Li_2S$ | Cubic | 300K | |
| LiI | Cubic | 300K | |
| $P_2S_5$ | Triclinic | 300K | |

**Table S2.** Molecule number of $Li_2S$, $B_2S_3$, LiI and $P_2S_5$ in different composition of LSPBI solid electrolytes for self-diffusion coefficient calculation. Here, the ratio of $Li_2S$-$B_2S_3$-LiI-$P_2S_5$ follows the experimental study of 30$Li_2S$-25$B_2S_3$-45LiI-$a$$P_2S_5$, where $a$ varies from 0-10, corresponding to the LSPBI-*a*0, LSPBI-*a*3, LSPBI-*a*5, LSPBI-*a*10 composition, respectively. The initial structure is generated by placing the $Li_2S$, $B_2S_3$, LiI and $P_2S_5$ molecules into a simulation box by using Packmol.

| Composition | $Li_2S$ | $B_2S_3$ | LiI | $P_2S_5$ | Total atoms number (Dimension of simulation cell after melt-quench) |
|---|---|---|---|---|---|
| LSPBI-*a*0_diff | 1800 | 1500 | 2700 | 0 | 18300 (74.5 Å · 74.5 Å · 74.5 Å) |
| LSPBI-*a*3_diff | 1800 | 1500 | 2700 | 180 | 19560 (75.6 Å · 75.5 Å · 75.5 Å) |
| LSPBI-*a*5_diff | 1800 | 1500 | 2700 | 300 | 20400 (74.5 Å · 74.5 Å · 74.5 Å) |
| LSPBI-*a*10_diff | 1632 | 1360 | 2448 | 544 | 20400 (77.1 Å · 77.1 Å · 77.1 Å) |

**Table S3.** Molecule number of $Li_2S$, $B_2S_3$, LiI and $P_2S_5$ in different composition of LSPBI solid electrolytes for mechanical tensile test simulations. The ratio of $Li_2S$ - $B_2S_3$ - LiI - $P_2S_5$ follows the experimental study of $30Li_2S$ - $25B_2S_3$ - 45LiI - $aP_2S_5$, where $a$ varies from 0-100, corresponding to the LSPBI-$a$0, LSPBI-$a$5, LSPBI-$a$10, LSPBI-$a$25, LSPBI-$a$50, LSPBI-$a$75, LSPBI-$a$100 composition, respectively. The initial structure is generated by placing the $Li_2S$, $B_2S_3$, LiI and $P_2S_5$ molecules into a simulation box by using Packmol.

| Composition | $Li_2S$ | $B_2S_3$ | LiI | $P_2S_5$ | Total atoms number (Dimension of simulation cell after melt-quench) |
|---|---|---|---|---|---|
| LSPBI-$a$0_mech | 1980 | 1650 | 2970 | 0 | 20130 (50.9 Å · 50.9 Å · 184.7 Å) |
| LSPBI-$a$5_mech | 1800 | 1500 | 2700 | 300 | 20400 (51.7 Å · 51.8 Å · 187.8 Å) |
| LSPBI-$a$10_mech | 1632 | 1360 | 2448 | 544 | 20400 (50.5 Å · 50.5 Å · 202.1 Å) |
| LSPBI-$a$25_mech | 1248 | 1040 | 1872 | 1040 | 19968 (55.5 Å · 55.5 Å · 170.7 Å) |
| LSPBI-$a$50_mech | 918 | 765 | 1377 | 1530 | 20043 (56.8 Å · 56.8 Å · 174.6 Å) |
| LSPBI-$a$75_mech | 720 | 600 | 1080 | 1800 | 19920 (57.5 Å · 57.5 Å · 176.7 Å) |
| LSPBI-$a$100_mech | 600 | 500 | 900 | 2000 | 20100 (58.1 Å · 58.1 Å · 178.9 Å) |